\documentclass[longauth]{aa}

\usepackage{graphicx}
\usepackage{txfonts}
\usepackage{natbib}
\bibpunct{(}{)}{;}{a}{}{,}

\newcommand{\hess}{H.E.S.S.}
\newcommand{\rxj}{\object{RX~J1713.7$-$3946}}
\begin{document}
  
  \title{A detailed spectral and morphological study of the gamma-ray
    supernova remnant \object{RX~J1713.7$-$3946} with \hess}

  \author{F. Aharonian\inst{1}
    \and A.G.~Akhperjanian \inst{2}
    \and A.R.~Bazer-Bachi \inst{3}
    \and M.~Beilicke \inst{4}
    \and W.~Benbow \inst{1}
    \and D.~Berge \inst{1}
    \and K.~Bernl\"ohr \inst{1,5}
    \and C.~Boisson \inst{6}
    \and O.~Bolz \inst{1}
    \and V.~Borrel \inst{3}
% ...
}

  \offprints{D. Berge, \email{David.Berge@mpi-hd.mpg.de}}

  \institute{
    Max-Planck-Institut f\"ur Kernphysik, P.O. Box 103980, D 69029
    Heidelberg, Germany
    \and
    Yerevan Physics Institute, 2 Alikhanian Brothers St., 375036 Yerevan,
    Armenia
    \and
    Centre d'Etude Spatiale des Rayonnements, CNRS/UPS, 9 av. du Colonel Roche, BP
    4346, F-31029 Toulouse Cedex 4, France
    \and
    Universit\"at Hamburg, Institut f\"ur Experimentalphysik, Luruper Chaussee
    149, D 22761 Hamburg, Germany
    \and
    Institut f\"ur Physik, Humboldt-Universit\"at zu Berlin, Newtonstr. 15,
    D 12489 Berlin, Germany
    \and
    LUTH, UMR 8102 du CNRS, Observatoire de Paris, Section de Meudon, F-92195 Meudon Cedex,
    France
% ...
}

  \date{Received xxx September 2005 / Accepted xxx November 2005}

  \abstract
      {}
      {We present results from deep observations of the Galactic
    shell-type supernova remnant (SNR) \rxj\ (also known as
    \object{G347.3$-$0.5}) conducted with the complete \hess\ array in
    2004.}
      {Detailed morphological and spatially resolved spectral
    studies reveal the very-high-energy (VHE -- Energies $E >
    100$~GeV) gamma-ray aspects of this object with unprecedented
    precision. Since this is the first in-depth analysis of an
    extended VHE gamma-ray source, we present a thorough discussion of
    our methodology and investigations of possible sources of
    systematic errors.}
      {Gamma rays are detected throughout the whole SNR. The emission
    is found to resemble a shell structure with increased fluxes from
    the western and northwestern parts. The differential gamma-ray
    spectrum of the whole SNR is measured over more than two orders of
    magnitude, from 190~GeV to 40~TeV, and is rather hard with
    indications for a deviation from a pure power law at high
    energies. Spectra have also been determined for spatially
    separated regions of \rxj. The flux values vary by more than a
    factor of two, but no significant change in spectral shape is
    found. There is a striking correlation between the X-ray and the
    gamma-ray image. Radial profiles in both wavelength regimes reveal
    the same shape almost everywhere in the region of the SNR.}
      {The VHE gamma-ray emission of \rxj\ is phenomenologically
    discussed for two scenarios, one where the gamma rays are produced
    by VHE electrons via Inverse Compton scattering and one where the
    gamma rays are due to neutral pion decay from proton-proton
    interactions. In conjunction with multi-wavelength considerations,
    the latter case is favoured. However, no decisive conclusions can
    yet be drawn regarding the parent particle population dominantly
    responsible for the gamma-ray emission from \rxj.}

  \authorrunning{F. Aharonian et al.}
  \titlerunning{The $\gamma$-ray supernova remnant
  \object{RX~J1713.7$-$3946}}
  
  
  \maketitle

  \section{Introduction}
  It is commonly believed that the only sources capable of supplying
  enough energy output to power the flux of Galactic cosmic rays are
  supernova explosions~\citep[e.g.,][]{Ginzburg}. At the present time
  there are two main arguments for this hypothesis: firstly, estimates
  of the power required to sustain the observed nuclear Galactic
  cosmic-ray population show that about 10\% of the mechanical energy
  released by the population of Galactic supernovae would suffice, or,
  in other words, that supernova remnants could be the sources of the
  Galactic cosmic rays if the average acceleration efficiency in a
  remnant is about 10\%. Secondly, a rather well developed theoretical
  framework for the acceleration mechanism, diffusive shock
  acceleration~\citep[for reviews see
  eg,][]{Blandford,Jones,MalkovDrury}, exists and it indeed predicts
  acceleration efficiencies in excess of 10\%.

  The best way of proving unequivocally the existence of
  very-high-energy (VHE) particles, electrons or hadrons, in the
  shells of supernova remnants (SNRs) is the detection of VHE (about
  100~GeV up to a few tens of TeV) gamma rays produced either via
  Inverse Compton (IC) scattering of VHE electrons off ambient photons
  or in interactions of nucleonic cosmic rays with ambient matter. As
  was argued already in~\citet{DAV}, a system of imaging atmospheric
  Cherenkov telescopes with a large field of view provides the most
  powerful measurement technique for extended nearby SNRs at these
  very high energies. One should note that there exist two other
  experimental approaches to trace VHE cosmic rays, the detection of
  X-rays, which suggests the presence of VHE
  electrons~\citep{Koyama95}, and of high-energy neutrinos, which
  probe exclusively nuclear particles.

  A prime candidate for gamma-ray observations is the SNR \rxj, in
  particular because of its close association with dense molecular
  clouds along the line of sight~\citep{FukuiUchiyama,Moriguchi},
  which might suggest a scenario of a supernova shell overtaking dense
  molecular clouds, leading to a detectable VHE gamma-ray signal from
  hadronic interactions, as described in~\citet{ADV}. \rxj, situated
  in the Galactic plane, constellation Scorpius, was discovered in
  soft X-rays in 1996 in the ROSAT all-sky
  survey~\citep{Pfeffermann}. It is roughly $70\arcmin$ in diameter
  and exhibits bright X-ray emission dominantly from its western
  shell. ASCA observations revealed that the X-ray emission is a pure
  non-thermal continuum without detectable line
  emission~\citep{Koyama,Slane}. X-ray observations have also been
  conducted with Chandra and XMM with their superior angular
  resolution. Chandra observed a small region in the bright
  northwestern part of the
  SNR~\citep{UchiyamaChandra,Lazendic}. Despite distinct brightness
  variations within this small field, the corresponding X-ray spectra
  were all found to be well described by power-law models with similar
  absorbing column densities and photon indices, albeit with rather
  large statistical uncertainties. XMM covered the remnant almost
  completely in five pointings~\citep{CassamXMM,HiragaXMM}. Also on
  this much larger scale, a highly inhomogeneous and complex
  morphology was found in the western part of the SNR with two narrow
  rims resembling a double-shell structure running from north to
  south. The (non-thermal) X-ray spectra, when fit with a power law,
  exhibit strong variations in photon index across the remnant $(1.8 <
  \Gamma < 2.6)$ and the hydrogen column density $N_{\mathrm{H}}$ was
  found to vary significantly $(0.4\times 10^{22}\ \mathrm{cm}^{-2}
  \leq N_{\mathrm{H}} \leq 1.1\times 10^{22}\ \mathrm{cm}^{-2})$. The
  spectra of the central and the western parts differ clearly at low
  energies, possibly indicating an increase in column density of
  $\Delta N_{\mathrm{H}} \approx 0.4\times 10^{22}\ \mathrm{cm}^{-2}$
  towards the west. Furthermore, a positive correlation between X-ray
  brightness and absorption was interpreted as being due to the shock
  front of \rxj\ impacting a molecular cloud in the west which was
  assumed to be responsible for the absorption. Further support for
  this scenario is lent by CO line emission observations with the
  NANTEN telescope~\citep{FukuiUchiyama,Moriguchi}, which suggest that
  the SNR is interacting with molecular clouds in this region at a
  distance of 1~kpc from the Solar System. The non-thermal X-ray
  emission is possibly associated with interactions between the cloud
  and the western part of the SNR shell.

  Age and distance of the SNR are under debate and have been revised
  quite a few times. Initially, \citet{Koyama} had derived a distance
  of 1~kpc and correspondingly an age of about 1000 years from the
  column density towards the source as estimated from ASCA X-ray
  observations. \citet{Slane} on the other hand have derived a larger
  distance of 6~kpc (corresponding to an age of about 10\,000 years)
  based on the possible association of \rxj\ with a molecular cloud in
  this region and the \ion{H}{ii} region \object{G347.6+0.2} to its
  northwest. Both the latest XMM and NANTEN findings are consistent
  with the remnant being closer, at 1~kpc, which might support the
  hypothesis of \citet{Wang}, that \rxj\ is the remnant of a AD393
  guest star which, according to historical records, appeared in the
  tail of constellation Scorpius, close to the actual position of
  \rxj. The high surface brightness of this object, both in VHE gamma
  rays and non-thermal X-rays, suggests that it is close to the
  evolutionary phase where the shocks are most powerful. While hardly
  a conclusive argument, this implies that the remnant is observed at
  the sweep-up time when the ejecta are interacting with approximately
  their own mass of swept-up ambient material and the energy flux
  through the shocks (both forward and reverse) peaks. Normally this
  would be at an age of a few hundred to a thousand years, which
  indeed supports the closer distance estimate.

  The radio emission of \rxj\ is very faint~\citep{Lazendic} which
  puts it into a peculiar class of shell-type SNRs with dominantly
  non-thermal X-ray and only very faint radio emission. The only other
  known object of this type is \object{RX~J0852.0$-$4622}
  (\object{G266.2$-$1.2})~\citep{AschenbachVelaJr,SlaneVelaJr}.

  \rxj\ was detected in VHE gamma rays by the CANGAROO collaboration
  in 1998~\citep{CANGI} and re-observed by CANGAROO-II in 2000 and
  2001~\citep{CANGII}. Recently \hess, a new array of imaging
  atmospheric Cherenkov telescopes operating in Namibia, has confirmed
  the detection~\citep{Hess1713}. This was the first independent
  confirmation of VHE gamma-ray emission from an SNR
  shell. Furthermore, the \hess\ measurement provided the first ever
  resolved gamma-ray image at very high energies. The complex
  morphology of \rxj\ was clearly unraveled. Together with the \hess\
  detection of \object{RX~J0852.0$-$4622}~\citep{HessVelaJr} there are
  currently two spatially resolved VHE gamma-ray SNRs with a
  shell-like structure which agrees well with that seen in
  X-rays. These two objects may well be the brightest SNRs in the VHE
  gamma-ray domain in the whole sky; anything equally bright in the
  Northern sky would have been clearly seen in the Milagro
  survey~\citep{MilagroAllSky}, and the \hess\ Galactic plane
  survey~\citep{HessPlaneScan} reveals no SNRs brighter than \rxj\ or
  \object{RX~J0852.0$-$4622} in the region covered.

  The interpretation of the gamma-ray emission mechanisms for \rxj\
  has been the subject of debate. From their flux level, the CANGAROO
  collaboration interpreted in~\citet{CANGI} the gamma rays as IC
  emission, whereas in \citet{CANGII}, after re-observations with
  CANGAROO-II, neutral pion decay was put forward as an explanation
  instead. The proposed model was then heavily disputed by
  \citet{ReimerPohl} and \citet{Butt} because of its conflict in the
  GeV regime with the flux of the nearby EGRET source
  \object{3EG~1714$-$3857}~\citep{Hartmann}. Further attempts to model
  the broadband spectrum of \rxj\ were
  undertaken~\citep[e.g.,][]{Ellison,UchiyamaChandra,Pannuti,Lazendic}. However,
  they did not result in unequivocal conclusions concerning the
  acceleration mechanisms of the highest-energy particles or the
  origin of the VHE gamma rays from this source.
  
  Here we report on follow-up observations of \rxj\ with the complete
  \hess\ telescope array, conducted in 2004. The large field of view
  together with the high sensitivity of the system enable us to
  undertake for the first time detailed morphological studies in VHE
  gamma rays and measure spectral parameters in different regions of
  the SNR.

  The paper is organised as follows: In Sect.~\ref{sec:DataProcessing}
  we present the data set and illustrate the performance of \hess\ for
  observations of extended gamma-ray sources. We explain in detail the
  analysis methods applied here in order to extract images and
  spectra. In Sect.~\ref{sec:Results} the results of spectral and
  morphological studies are presented along with systematic tests that
  have been performed in order to assure the validity of the
  analysis. Section~\ref{sec:Multiwavelength} presents
  multi-wavelength data of \rxj\ and its surroundings aiming at
  putting our measurement at TeV energies into the context of the
  available data as preparation for a broadband modelling of the
  spectral energy distribution (SED) in Sect.~\ref{sec:Modelling}. We
  discuss two scenarios for the generation of VHE gamma rays, a purely
  electronic and a purely hadronic one. The results are finally
  summarised in Sect.~\ref{sec:Summary}.

  \section{Data processing}\label{sec:DataProcessing}
  \subsection{\hess\ observations}
  Observations of \rxj\ were conducted between April and July 2004
  with the High Energy Stereoscopic System (\hess), a system of four
  imaging atmospheric Cherenkov
  telescopes~\citep{JimHessStatus,HofmannHessStatus} situated in the
  Khomas Highland of Namibia, at $23\degr16\arcmin\
  \mathrm{S}$~$16\degr 30\arcmin\ \mathrm{E}$, 1800~m above sea
  level. Each of the 13-m-diameter
  telescopes~\citep{BernloehrHessOptics,CornilsII} has a tessellated
  Davies-Cotton mirror of $107~\mathrm{m}^2$ area and is equipped with
  a 960-photomultiplier-tube camera ~\citep{PascalHessCamera} covering
  a large field of view of $5\degr$ diameter. During stereoscopic
  observations, an array-level hardware trigger requires each shower
  to be detected by at least two telescopes simultaneously allowing
  for efficient suppression of the vast number of hadronic and muonic
  background events ~\citep{StefanHessTrigger}. The point source
  sensitivity reaches $1\%$ of the flux of the Crab nebula for long
  exposures ($\approx 25$~hours).

  \begin{figure}
    \resizebox{\hsize}{!}{\includegraphics[draft=false]{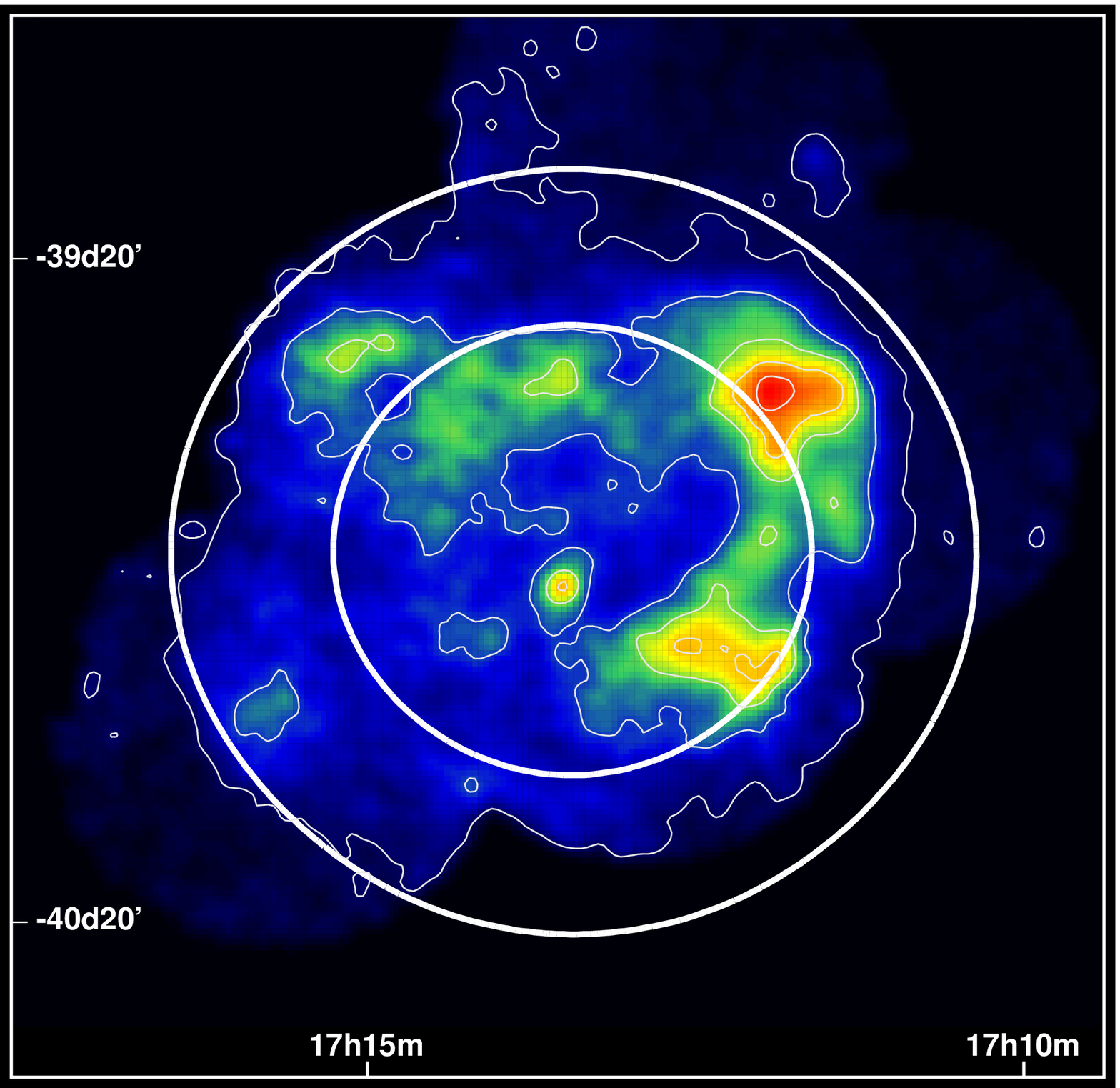}}
    \caption{X-ray image of \rxj\ (colour scale and thin grey contour
      lines, 1-3~keV, from \citet{AscaI}). The superimposed thick white
      contours indicate the $94\%$ and $98\%$ levels of the
      detection-efficiency weighted \hess\ exposure, given by the
      product of relative detection efficiency and the observation
      time. It can be seen that the relative gamma-ray detection
      efficiency between the centre region and the edges of the SNR
      differs only by $\approx 5\%$.}
    \label{fig:Exposure}
  \end{figure}
  The observations were mostly performed in \textit{wobble mode}
  around the SNR centre. In this mode the telescopes were positioned
  such that the centre of the SNR was offset $\pm 0.7\degr$ in
  declination or right ascension away from the pointing direction of
  the telescope system, changing to the next position every $28$
  minutes. Towards the end of the observation campaign, pure on-source
  pointings in which the centre of the SNR was coincident with the
  system centre were additionally performed. In each of the five
  pointings the SNR \rxj, roughly $1\degr$ in diameter, was fully
  contained in the $\approx 5\degr$ field of view of the system. The
  resulting effective exposure distribution (the product of the
  detection efficiency and the exposure) is illustrated in
  Fig.~\ref{fig:Exposure}, where the ASCA X-ray
  measurement~\citep{AscaI} is shown (colour scale and thin contour
  lines) with superimposed white, almost circular contours indicating
  the $94\%$ and $98\%$ levels of the effective exposure. The
  observation strategy for this data set combined with the detector
  efficiency results in a very flat plateau in the region of the SNR;
  from the centre to the boundaries the relative gamma-ray detection
  efficiency decreases by only about $5\%$, which is a great advantage
  compared to ASCA, for example. Not only is the SNR fully contained
  in all of the five pointing positions, but one can also disregard
  for most purposes the modest change in relative detection efficiency
  from one region of the SNR to another.
  
  \subsection{Data sample}
  The data comprise a total exposure time of $40$~hours. Rejecting
  data taken under bad weather conditions, 36~hours of observation
  time corresponding to 33~hours of live time remain for the
  analysis. The zenith angle of observations ranged from
  $16\degr$~to~$56\degr$ with a mean of $26\degr$; it should be noted
  that about $68\%$ of the data were taken at small zenith angles
  between $16\degr$~to~$26\degr$. The energy threshold (defined by the
  peak gamma-ray detection rate for a given source spectrum after all
  gamma-ray selection cuts) of the system increases with zenith
  angle. For the observations presented here, assuming a spectrum
  appropriate for \rxj, the threshold was $\approx 180$~GeV at
  $16\degr$, $\approx 340$~GeV at $40\degr$, and $\approx 840$~GeV at
  $56\degr$.

  \subsection{Data preprocessing}
  The analysis technique applied here is described in detail in
  \citet{HessPKS2155}. After calibration of the
  data~\citep{HessCalib}, tail-cuts image cleaning is applied and the
  shower images in each telescope are parametrised in terms of their
  centre of gravity and second moments~\citep{Hillas}. Stereoscopic
  event reconstruction based on the intersection of image axes yields
  the shower direction, providing a resolution of $\approx 0.1\degr$
  for individual gamma rays. Given the geometry of the shower, cuts
  (optimised on Monte Carlo gamma-ray simulations and \textit{OFF}
  source data, i.e. data without gamma-ray signal) are applied to
  select gamma-ray candidates and to suppress the vast hadronic
  background. The gamma-ray energy is estimated from the image
  intensity and the reconstructed shower geometry yielding a
  resolution of $\approx 15\%$. There is one main difference between a
  point-source and an extended-source analysis. In the latter case,
  the cut on the squared distance of events to the assumed source
  location is greatly increased in order to reflect the large source
  extension. In that case the energy resolution worsens slightly (to
  $\la 20\%$ for a source of the size of \rxj) since an increased
  number of badly reconstructed events are included in the analysis.

  \begin{figure*}
    \centering
    \includegraphics[width=17cm]{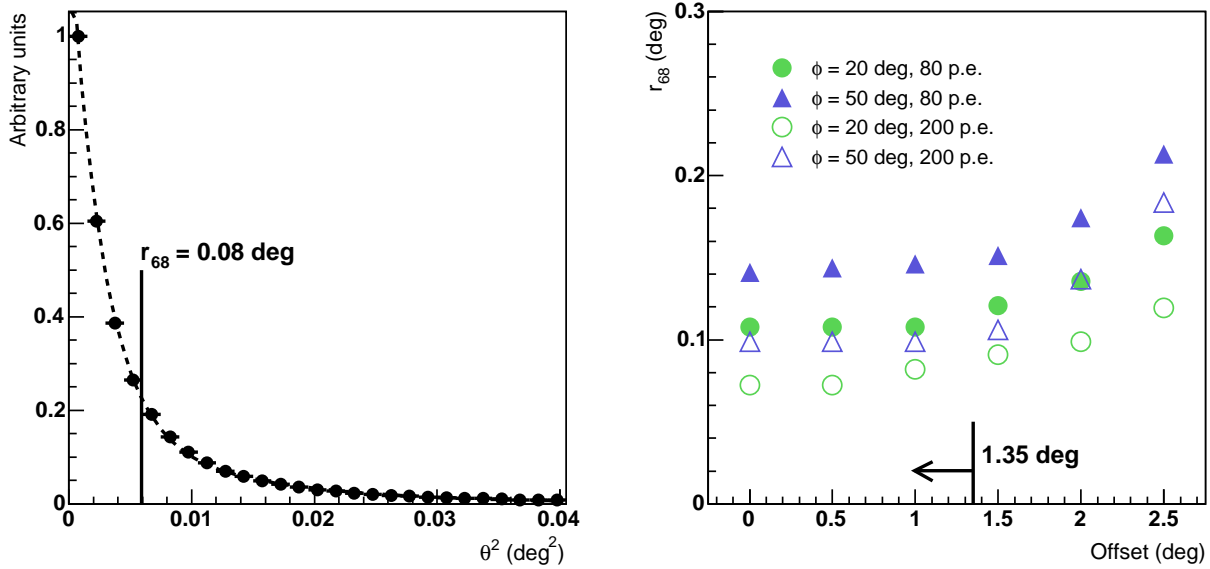}
    \caption{\textbf{Left hand side:} Squared angular distance between
    reconstructed and true direction for a Monte Carlo gamma-ray point
    source (\textit{point-spread function}). A cut on the minimum size
    of images of 200 photo-electrons was applied. The simulated gamma
    rays followed a power law in energy with a photon index of 2. The
    zenith angle distribution and offset of the source with respect to
    the telescope optical axis was matched to the actual data set of
    \rxj. The solid circles are the Monte Carlo histogram, the fit of
    a double Gaussian (dashed line) describes the point-spread
    function reasonably well. Indicated is the $68\%$ containment
    radius of $0.08\degr$ (obtained from the fit), which is taken as
    the resolution of the data set. \textbf{Right hand side:} $68\%$
    containment radii of the point-spread function determined from
    simulations as function of the offset between the source direction
    and telescope pointing. Shown are simulations for $50\degr$
    (triangles) and $20\degr$ zenith angle (circles) for two different
    size cuts (80 and 200 photo-electrons). The maximum offset of the
    actual data set of $1.35\degr$, given by the wobble displacement
    of $0.7\degr$ and the approximate radius of the SNR of
    $0.65\degr$, is indicated by the arrow.}
    \label{fig:Psf}
  \end{figure*}

  \subsection{\hess\ performance for extended sources}\label{subsec:HessPerformance}

  The large field of view of the \hess\ telescope system of $\approx
  5\degr$ diameter provides reasonable sensitivity for point sources
  at an angular distance up to $2\degr$ from the pointing direction of
  the telescope system (the point-source off-axis sensitivity derived
  from Monte Carlo simulations has been confirmed via observations of
  the Crab nebula, see \citet{HessCrab}). Given the source diameter of
  up to $1.3\degr$ and the offsets of $0.7\degr$ between the centre of
  the SNR and the telescope pointing direction during observations, it
  is important that the gamma-ray point-spread function is well
  behaved and does not broaden significantly with increasing offset
  from the pointing direction. Figure~\ref{fig:Psf} (left) shows the
  squared angular difference $\theta^2$ between the reconstructed and
  the true direction of a simulated point source. The initial
  simulations have been generated at a number of fixed zenith angles
  between $0\degr$ and $70\degr$. Taking then the zenith angle
  distribution of the actual data set and forming from that the
  weighted sum of the simulated $\theta^2$ distributions one obtains a
  resolution (taken as the $68\%$ containment radius) of
  $0.08\degr$. This is an order of magnitude smaller than the source
  diameter, which implies that in terms of angular resolution \hess\
  is well suited for morphological studies of \rxj. The right hand
  side of Fig.~\ref{fig:Psf} illustrates the dependence of the angular
  resolution on the offset between source position and pointing
  direction for two representative zenith angles of $20\degr$ and
  $50\degr$, for two different cuts on the minimum camera image size
  of 80 and 200 photo-electrons. It can be seen that the application
  of a higher cut on the image size improves the direction
  reconstruction by about $20\%$ since only well defined camera images
  are used in the shower reconstruction, reducing fluctuation
  effects. However, the improved resolution is achieved at the expense
  of an increased energy threshold. In any case, for offsets smaller
  than $1.35\degr$, which is the maximum offset under which parts of
  the SNR were observed, the resolution changes only slowly with
  offset and worsens by less than $10\%$. It was furthermore
  demonstrated with observations of the point-like source
  \object{PKS~2155$-$304}~\citep{HessPKS2155}, that simulated
  point-spread functions agree well with data distributions and are
  thus well understood.
  
\begin{figure}
    \resizebox{\hsize}{!}{\includegraphics[draft=false]{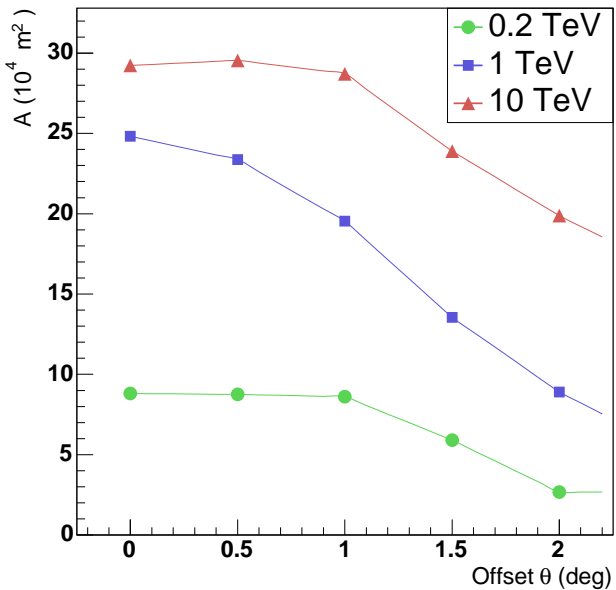}}
    \caption{Effective detection area $A$ for a moderate zenith angle
    of $20\degr$ as a function of offset angle $\theta$ between the
    source location and the pointing direction of the system. Plotted
    are values at three representative energy values of 0.2~TeV (near
    the post-cut trigger threshold), 1~TeV, and 10~TeV. The markers at
    $0\degr$, $0.5\degr$, $1\degr$, $1.5\degr$, and $2\degr$ represent
    the offset values for which simulations are available, in between
    two simulated offsets effective area values are interpolated
    linearly.}
    \label{fig:EffArea}
  \end{figure}

  The effective gamma-ray detection area depends on trigger conditions
  and analysis cuts. Well above the trigger threshold of the system,
  it is of the order of the area of the Cherenkov light pool (emitted
  by the secondary particle shower) on the ground. Typical effective
  area curves as a function of the offset angle between the gamma-ray
  source and the pointing direction of the system are shown in
  Fig.~\ref{fig:EffArea} for a zenith angle of $20\degr$. The
  effective areas were determined from Monte Carlo simulations of a
  gamma-ray point source, accounting for the large extension of \rxj\
  by increasing the cut on the maximum distance to the assumed source
  position.
  Since for an extended source like \rxj\ the flux has to be
  integrated over a larger solid angle than for point sources, the
  sensitivity of the system is reduced due to an increased background
  level. For moderate zenith angles and an integration region of
  $0.65\degr$ radius around a gamma-ray source at $1\degr$ offset from
  the pointing direction (matched to the data set described here),
  simulations reveal that, for a source like \rxj, the sensitivity
  drops roughly by a factor of four as compared to a point-source
  analysis.

  \subsection{Analysis details}\label{subsec:AnalysisTechnique}
  The following section explains briefly the analysis applied to
  extract gamma-ray images and spectra of \rxj\ from the data.
  \subsubsection{Morphology}\label{subsubsec:ImageAnalysis}

  \begin{figure}
    \resizebox{\hsize}{!}{\includegraphics{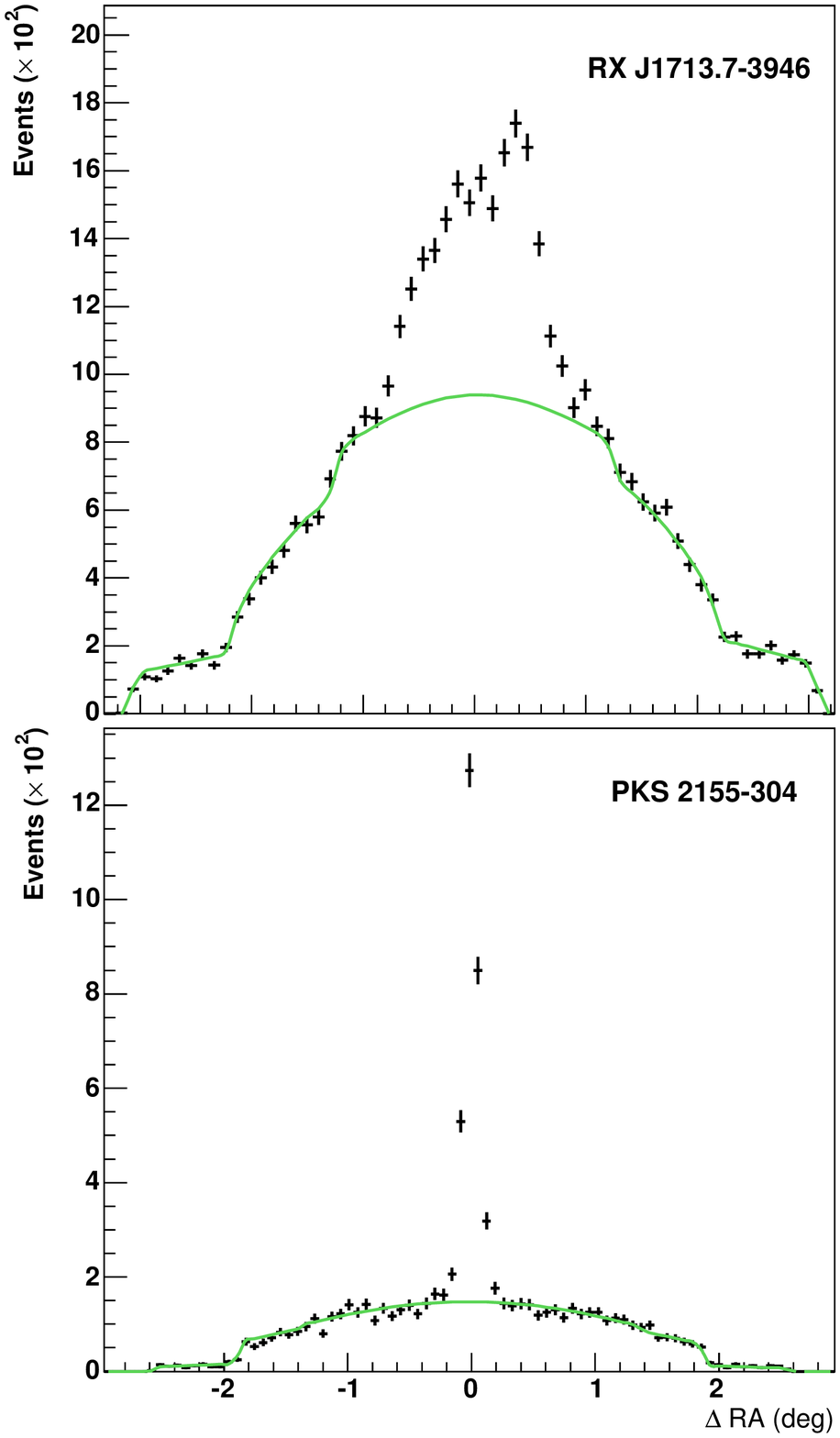}}
    \caption{As illustration of the good match between data and
      background model, the data set of \rxj\ (upper panel) is
      compared to the high statistics data set of
      \object{PKS~2155$-$304} (lower panel) which appears as a point
      source for \hess. Shown are $1.3\degr$ and $0.2\degr$ slices
      along the RA axis through the centre of \rxj\ and
      \object{PKS~2155$-$304}, respectively, for the data (black) and
      the background model (green). The good match both in shape and
      absolute level between the two curves illustrates the validity
      of the approach. The steps in the distributions (in the case of
      \rxj\ at $\pm 1.1\degr$ and $\pm 2.1\degr$) are artefacts of the
      analysis: the usable range in the field of view of every
      observation was restricted to a radius of $2\degr$ (out of
      $2.5\degr$ physical camera radius) around the camera centre to
      allow only for well reconstructed events with good camera
      acceptance, and the figure combines data from different
      pointings.}
    \label{fig:Slices}
  \end{figure}

  The gamma-ray images shown throughout the paper represent, unless
  otherwise stated, gamma-ray excess counts with background
  subtracted. For the generation of these images a cut of 200
  photo-electrons on the minimum camera image size in each telescope
  is applied to select a subset of events with superior angular
  resolution (see also Fig.~\ref{fig:Psf} and
  Sect.~\ref{subsec:HessPerformance}). The background was estimated
  using about 160 hours of \hess\ observations without any or only
  very faint gamma-ray sources in the field of view. All the events in
  these reference observations passing gamma-ray cuts are assumed to
  be gamma-ray like background events and are used to estimate the
  background for the given data set. For that purpose the set of OFF
  runs has been divided into distinct zenith angle bands to account
  for the dependance of the system's gamma-ray acceptance on
  observation altitude. Besides that, the acceptance depends only on
  the angular distance between shower and pointing direction and is to
  a very good approximation radially symmetric with respect to the
  pointing direction. Therefore, a radial 1-D lookup (number of
  background events as a function of squared distance to the pointing
  centre) can be used in each zenith angle band for the background
  estimation. Given an observation at a certain zenith angle, a model
  background is created by selecting the 1-D radial lookup from the
  zenith-angle band that matches the zenith angle of the
  observation. A 2-D background map of the sky is then created by
  rotating the corresponding 1-D lookup. Finally, the overall
  background map is created as the exposure weighted sum of the
  individual maps. A global normalisation factor $\alpha$ is
  calculated as ratio of the number of events in the data to the
  number of events in the background model, excluding regions that
  emit gamma rays. Figure~\ref{fig:Slices} illustrates the validity of
  the approach for two \hess\ data sets, \rxj\ (upper panel) and
  \object{PKS~2155$-$304} (lower panel), the latter being a point
  source for \hess. Shown are slices along right ascension through the
  source centres. Overlaid on both data curves are the normalised
  background models of the whole data sets. In both cases, at
  different regions in the sky, for an extended and a point-like
  gamma-ray source, there is clearly a good match between model and
  data in regions outside the gamma-ray sources.

  Images of the gamma-ray excess are obtained by subtracting the
  normalised background model from the data in each bin $i$ of the 2-D
  map:
  \begin{eqnarray*}
  N_{\mathrm{excess,i}} = N_{\mathrm{data,i}} - \alpha \,
  N_{\mathrm{background,i}} \, .
  \end{eqnarray*}
  Subsequently, these images are smoothed with a Gaussian to reduce
  statistical fluctuations. 

  Typically the standard deviation for the smoothing is matched to the
  resolution of the data set, namely $30\%$~-~$50\%$ of the $68\%$
  containment radius of the point-spread function. Here, for the data
  of \rxj, a smoothing radius of $2\arcmin$ is used. The resulting
  count maps are in units of integrated excess counts per Gaussian
  sigma of the smoothing function.

  No correction for the falloff of detection efficiency towards the
  edges of the field of view is applied because in the region of
  interest, around the SNR \rxj, the variation in gamma-ray acceptance
  is negligible (see Fig.~\ref{fig:Exposure}).

  \subsubsection{Spectral analysis}\label{subsec:SpectralAnalysis}
  \begin{figure}
    \resizebox{\hsize}{!}{\includegraphics{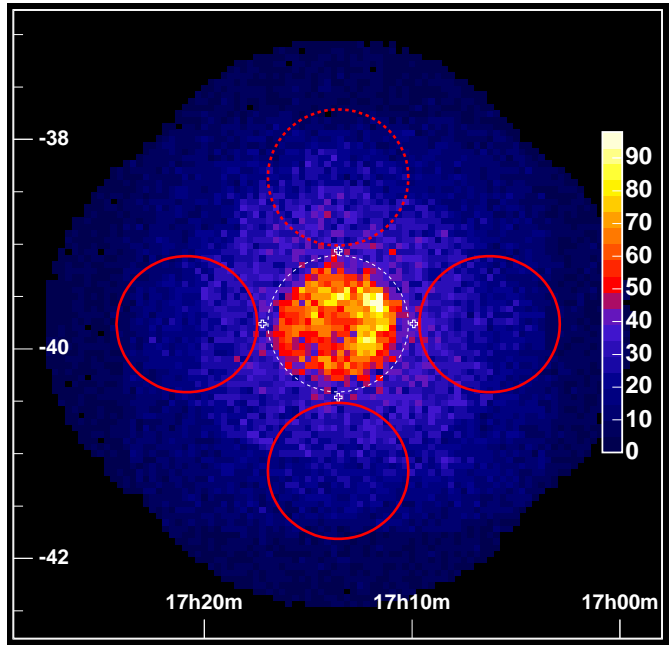}}
    \caption{Count map of gamma-ray candidates for the region around
    \rxj. A size cut of 200 photoelectrons on the camera images was
    applied. The bins are uncorrelated and the background is not
    subtracted. The white dashed circle indicates the region used to
    extract the spectrum of the whole SNR (the \textit{ON} region),
    the red circles indicate the OFF regions, the regions used for
    background estimation for the spectrum in each of the four wobble
    observation positions, which are marked as white crosses. The
    dashed red circle to the north of the SNR indicates an OFF region
    that was not used in the spectral analysis because it contains a
    gamma-ray source discovered in the \hess\ Galactic plane
    survey~\citep{HessPlaneScan}.}
    \label{fig:OffRegions}
  \end{figure}
  For the spectral analysis, the \hess\ standard cut on the image size
  of 80 photo-electrons is applied to the data. To obtain a spectrum
  of a certain region in the sky, all events with reconstructed
  direction in that particular region are considered as ON events. A
  complication arises for the background estimation. The gamma-ray
  acceptance and therefore the background level depend strongly on
  energy; one cannot, as for the image generation, simply use a 1-D
  radial lookup (which is integrated over all energies) as a
  background estimate. Instead, the acceptance lookups would have to
  be generated in energy bins which in practice is difficult to
  handle. Another approach was applied here for the spectral analysis:
  background (OFF) events were selected from the same field of view,
  from the same data run, by selecting regions of the same size and
  form as the ON region, but displaced on a circle around the pointing
  direction of the system. The circle is chosen such that the OFF
  regions are at exactly the same offset (that is, at the same angular
  distance to the pointing direction) as the ON region. A minimum
  distance between the ON and OFF regions of $0.1\degr$ is required to
  avoid gamma-ray contamination. Furthermore, known gamma-ray sources
  in the field of view not associated with the test region are
  excluded from the OFF regions. This approach ensures that background
  events are taken at the same zenith and offset angles, which is
  crucial because of the dependence of the effective detection areas
  upon these two quantities, and it uses more or less the same region
  of sky, with similar night-sky-background noise. For an object of
  the size of \rxj\ this results in one ON and OFF region, the latter
  being simply the reflection of the former at the system centre. This
  is illustrated in Fig.~\ref{fig:OffRegions}, where the OFF regions
  used for each observation position are drawn.

  After the geometrical selection, in order to obtain a differential
  gamma-ray flux $\mathrm{d}N / \mathrm{d}E$ in units of
  $(\mathrm{TeV}^{-1}~\mathrm{cm}^{-2}~\mathrm{s}^{-1})$, ON and OFF
  events ($N_{\mathrm{ON}}$ and $N_{\mathrm{OFF}}$) are binned
  logarithmically in energy and divided by the mean effective area and
  the exposure time in each bin. The energy dependent effective area
  is determined for each data run, for the corresponding zenith
  $(\theta_{\mathrm{z}})$ and offset $(\theta)$ angle range,
  multiplied by the live time of the run and added up. Then, for each
  energy bin $i$, the bin entries are divided by the width
  ($\Delta_{\mathrm{i}}$) of that bin to obtain a differential flux
  value. The differential flux results from subtracting the
  differential OFF- from the ON-flux histogram:
  \begin{eqnarray*}
    \left(\frac{\mathrm{d}N}{\mathrm{d}E}\right)_{\mathrm{i}} =
    \frac{N_{\mathrm{on,i}}}{\Delta_{\mathrm{i}} \, \Sigma_{\mathrm{runs}} T
    A_{\mathrm{i}} (\theta_{\mathrm{z}},\theta)} - \alpha_{\mathrm{i}} \,
    \frac{N_{\mathrm{OFF,i}}}{\Delta_{\mathrm{i}} \, \Sigma_{\mathrm{runs}}
    T A_{\mathrm{i}} (\theta_{\mathrm{z}},\theta)}
  \end{eqnarray*}
  The normalisation factor $\alpha_{\mathrm{i}}$ is determined from
  the ratio of the areas of the ON and OFF regions used during
  analysis of the whole data set and takes the exposure of the
  different observations into account.

  An alternative approach is the event-by-event usage of the effective
  area, which served as a systematic check. Rather than determining a
  mean effective area for the whole data set, each event is weighted
  with the inverse of the effective area, taking the event zenith
  angle, offset and energy.

  A complication arises from the dependence of the effective detection
  area on zenith angle and offset. The Monte Carlo effective areas are
  generated at certain discrete zenith angles and offsets. When
  generating the mean effective area corresponding to a certain zenith
  angle and offset range, the effective area is interpolated linearly
  between the simulation values (see Fig.~\ref{fig:EffArea}, where the
  markers indicate values available from simulations and the lines are
  the linear interpolations).

  \section{Results}\label{sec:Results}
  With the \hess\ data set, the morphology of \rxj\ and its spectrum
  are resolved with high precision. Given that this is the first in
  depth analysis of such an extended source in VHE gamma rays, we
  present in the following first selected examples of extensive
  systematic tests that were performed in order to assure the
  stability of the analysis and then discuss the results.

  \subsection{Morphology}\label{subsec:morphology}
  \begin{figure}
    \resizebox{\hsize}{!}{\includegraphics[draft=false]{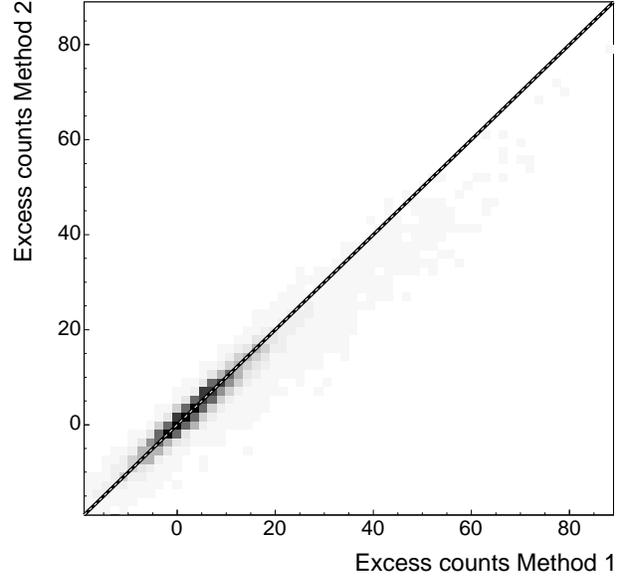}}
    \caption{Correlation between the standard calibration, event
    reconstruction, and background reduction method (\textit{Method
    1}) and an alternative, completely independent approach
    (\textit{Method 2}, see main text for more details).  The black
    solid and white dashed lines run through the origin with a slope
    of 1. A linear correlation is clearly visible, deviations from the
    line originate from different cut efficiencies of the two
    methods.}
    \label{fig:MorphologyCheck}
  \end{figure}  
  When analysing the morphology of an extended source, one aims for
  the best possible resolution with, at the same time, sufficient
  event statistics. In order to explore image structures and their
  stability, the data set was analysed using the same calibration and
  analysis software, but applying different sets of cuts (like
  accepting only three- and four-telescope events and events with
  image amplitudes larger than 80, 200, and 400 photo-electrons),
  which resulted in different resolutions and different event
  statistics. Another important issue is the appropriate modelling and
  subtraction of the background. As a systematic test, alternative
  background models have been applied and the results compared. First,
  a set of OFF runs, taken at the same zenith angles with very similar
  night-sky noise as the ON runs, has been used. Furthermore, the
  morphology was cross-checked using a completely independent
  calibration and analysis approach, not only for the reconstruction
  of the shower geometry but also for background estimation. Rather
  than using standard (Hillas) parameters for image parametrisation
  and reconstructing shower geometry based on these parameters, this
  approach is based on a 3-D modelling of Cherenkov photon emission
  during the shower development in the atmosphere assuming rotational
  symmetry, thereby predicting pixel amplitudes~\citep[for more
  details, see][]{ModelPaper}. A comparison of the two analysis
  methods is shown in Fig.~\ref{fig:MorphologyCheck}; plotted is the
  correlation of gamma-ray excess counts for the sky region around
  \rxj. A linear correlation is clearly apparent illustrating the good
  agreement of the two independent methods.

  With the systematic tests mentioned above it could be shown that the
  main features of the gamma-ray morphology are stable when analysed
  with different cuts, different background models as well as with
  independently determined calibration coefficients and alternative
  analysis methods.

  \begin{figure}
    \resizebox{\hsize}{!}{\includegraphics[draft=false]{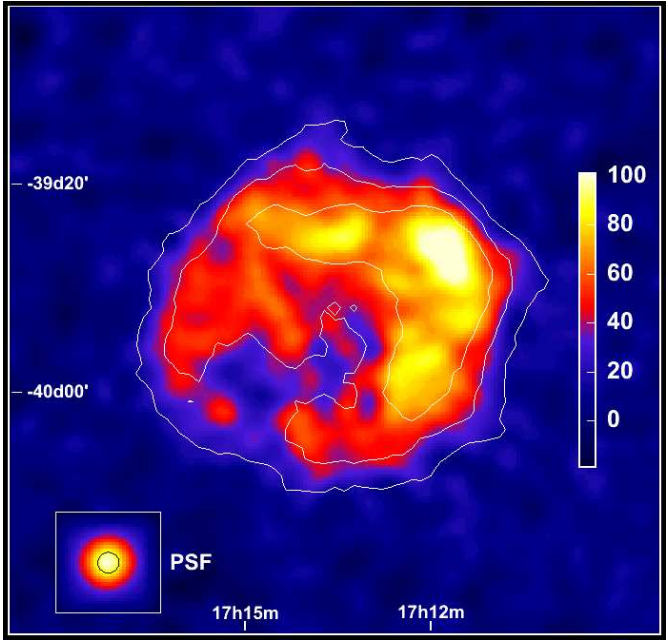}}
    \caption{Gamma-ray image of \rxj.  The linear colour scale is in
    units of excess counts (see Sect.~\ref{subsubsec:ImageAnalysis}
    for a description of image generation). The white contour lines
    indicate the significance of the different features, the levels
    are linearly spaced and correspond to 5, 10, and 15$\sigma$,
    respectively. The significance of each point has been calculated
    assuming a point source at that position, integrating events
    within a circle of $0.1\degr$ radius. In the lower left hand
    corner a simulated point source is shown as it would appear in
    this particular data set (taking the point-spread function and the
    smoothing into account) along with a black circle of $2\arcmin$
    radius denoting the $\sigma$ of the Gaussian the image is smoothed
    with.}
    \label{fig:Size200Image}
  \end{figure}  
  Figure~\ref{fig:Size200Image} shows a $2\degr \times 2\degr$ field
  of view around \rxj. A cut on the image size at 200 photo-electrons
  was applied resulting in a superior resolution of $0.08\degr$ (see
  Fig.~\ref{fig:Psf}). The corresponding energy range is $\approx
  300$~GeV to $\approx 40$~TeV. This image of \rxj\ confirms with much
  higher statistics the 2003 \hess\ measurement, shown for example in
  Fig.~1 of \citet{Hess1713}. There is no evidence for time
  variability, as expected for an object of the size of \rxj. The
  overall gamma-ray appearance resembles a shell morphology with
  bright emission regions in the western and northwestern part where
  the SNR is believed to impact molecular
  clouds~\citep{FukuiUchiyama,Moriguchi}. It is worth noting that
  there is a possible gamma-ray void in the central-southeastern
  region. The cumulative significance for the whole SNR is about
  $39\sigma$ with these hard cuts, which corresponds to an excess of
  $\approx 7700$ events from the region of \rxj. Drawn as white lines
  in Fig.~\ref{fig:Size200Image} are in addition the contours of
  significance of the gamma-ray signal (levels correspond to 5, 10,
  and 15$\sigma$). The significance has been calculated considering
  events that fall within an angle of $0.1\degr$ of each trial source
  position. Thus, the contours quantify the significance for each
  point as if there was a point source at that position. The
  background estimate was derived from OFF runs as described in
  Sect.~\ref{subsubsec:ImageAnalysis}. The brightest parts of the SNR
  exceed $20\sigma$. Except for the void structure, where the
  significance just exceeds $5\sigma$, most of the remaining emission
  regions are well above $10\sigma$.

  \begin{figure}
    \resizebox{\hsize}{!}{\includegraphics[draft=false]{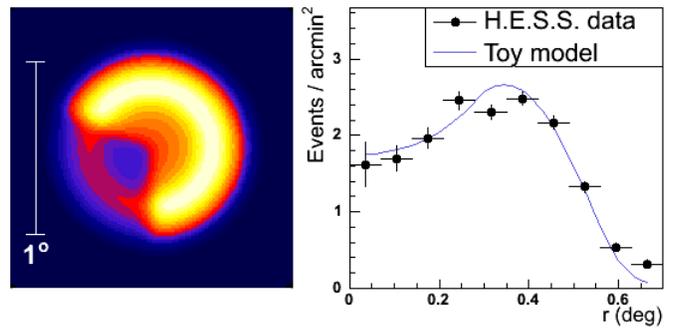}}
    \caption{Shown is a simple geometrical model for the emission from
    a thick sphere matched to the dimensions and relative fluxes of
    \rxj. \textbf{Left:} 2-D projection of a thick and spherical
    radiating shell, $1\degr$ in diameter, smoothed with the \hess\
    point-spread function. Adapted empirically to match the radial
    shape of the \hess\ data set, the dimensions of the geometrical
    sphere are 5.5~pc for the inner and 10~pc for the outer boundary
    if one assumes a distance of 1~kpc to the source, the presumed
    distance to \rxj. The emissivity in the northern, western, and
    southwestern part is a factor of two higher than in the southeast
    and east. \textbf{Right:} Radial profile from the geometrical
    model compared to the \hess\ data profile of \rxj. The centre
    coordinates used for the data plot are $\alpha_{\mathrm{J2000}} \, =
    \, \mathrm{17h13m33.6s}$, $\delta_{\mathrm{J2000}} \, = \, -39\degr
    45\arcmin 36\arcsec$. The geometrical model profile has been
    scaled to the same area as the data profile.}
    \label{fig:ToyShell}
  \end{figure}  
  From the gamma-ray image presented here it is clear that the
  emission regions cannot be distributed homogeneously in the
  \textit{sphere} \rxj. The image is neither rotational symmetric nor
  does it exhibit a shallow peak towards the centre. Instead, a shell
  seems to be apparent in the northern, and western to southwestern
  part. Apart from that, there is more or less uniform emission found
  in the rest of the SNR with a slight flux increase towards the
  southeastern boundary. This resembles very much the image one would
  expect from a thick spherical shell radiating gamma rays with
  enhanced emission on one side, as is illustrated in
  Fig.~\ref{fig:ToyShell} where a geometrical model of a thick
  radiating sphere is presented. The good match in shape between the
  data and the toy-model profile lends support to the assumption that
  indeed it is the shell of \rxj\ which radiates gamma rays.

  \begin{figure*}
    \centering
    \includegraphics[draft=false,width=17cm]{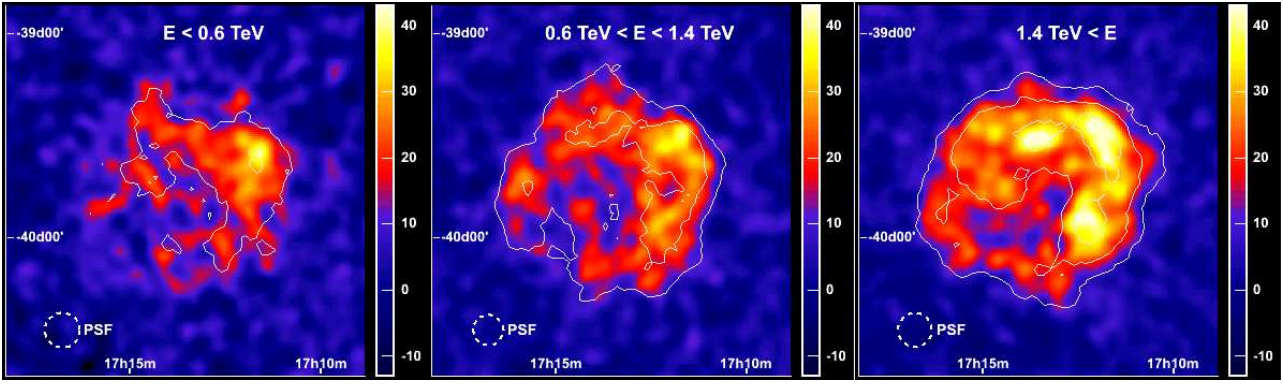}
    \caption{Morphology of \rxj\ as it appears at different
      energies. Shown from left to right are gamma-ray excess images
      with energies of $E~<~0.6~\mathrm{TeV}$,
      $0.6~\mathrm{TeV}~<~E~<~1.4~\mathrm{TeV}$, and
      $1.4~\mathrm{TeV}~<~E$. Drawn additionally as white lines are
      contours of significance, linearly spaced at 5, 10, 15$\sigma$
      (as in Fig.~\ref{fig:Size200Image}). Note the increase in the
      signal-to-noise ratio with increasing energy. The energy bands
      were chosen such that each band represents about a third of the
      full data set (taking events after cuts). Furthermore, all three
      images were smoothed with a Gaussian of $2 \arcmin$, which makes
      them directly comparable to each other, and to
      Fig.~\ref{fig:Size200Image}. The resolution in each energy band
      is indicated in the lower left hand corner of the images; the
      three data subsets have comparable resolutions of $\approx
      0.08\degr$ (the resolution of the intermediate energy band is
      about $6\%$ better). This might be counter-intuitive, given that
      at larger energies camera images get bigger and fluctuation
      effects become negligible thereby improving the energy and
      direction resolution. However, in this case that effect is
      compensated by the increasing mean zenith angle of the
      large-energy events.}
    \label{fig:EnergyBands}
  \end{figure*}
  Figure~\ref{fig:EnergyBands} shows three images of \rxj\ in three
  distinct energy bands, $E~<~0.6~\mathrm{TeV}$,
  $0.6~\mathrm{TeV}~<~E~<~1.4~\mathrm{TeV}$, and
  $1.4~\mathrm{TeV}~<~E$ (left to right). The energy ranges were
  chosen such that each band represents a third of the data set. Note
  that the angular resolution of all three images is roughly the same
  which makes them readily comparable. The signal-to-noise ratio of
  the low-energy image is evidently smaller. The shell-like morphology
  of the SNR is slightly blurred by fluctuations. Correspondingly, the
  significance contours indicate that only the bright northwestern
  half is significant in this energy band. In contrast, the whole
  remnant sticks out significantly in the two higher-energy
  bands. Most of the northwestern parts exceed $10\sigma$, the
  brightest spots even exceed $15\sigma$ for energies beyond 1.4~TeV.

\begin{figure}
    \resizebox{\hsize}{!}{\includegraphics[draft=false]{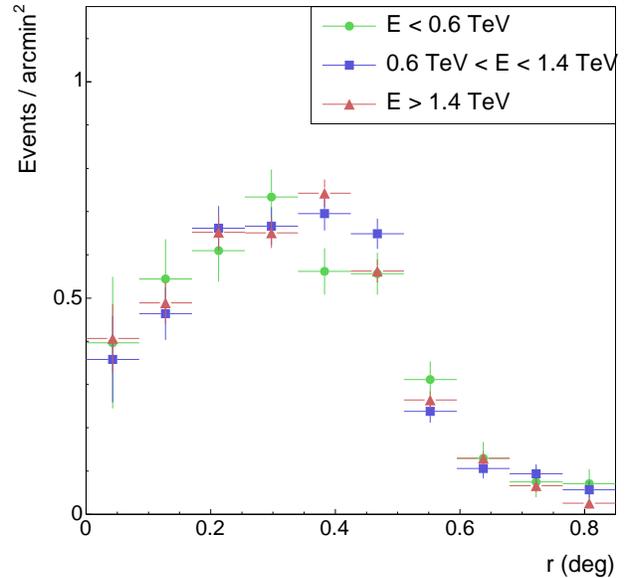}}
    \caption{Radial profiles around the centre of the SNR
  ($\alpha_{\mathrm{J2000}} \, = \, \mathrm{17h13m33.6s}$,
  $\delta_{\mathrm{J2000}} \, = \, -39\degr 45\arcmin 36\arcsec$) in
  the three energy bands plotted in Fig.~\ref{fig:EnergyBands},
  generated from the raw (that is, not smoothed) gamma-ray excess
  images. The intermediate and high-energy band images have improved
  signal-to-noise ratios. The radial profiles from these energy bands
  have been scaled such that they have the same area as the low-energy
  profile (scaling factors are $\approx 0.87$ and $\approx 0.65$,
  respectively), to enable direct comparisons.}
    \label{fig:EnergyBands_RadialProfiles}
  \end{figure}

  From the visual impression the remnant does seem to emit gamma rays
  more uniformly with increasing energy. However, within errors, the
  radial shape appears to be the same in all three energy bands, as
  seen from the radial excess profiles around the centre of the SNR
  shown in Fig.~\ref{fig:EnergyBands_RadialProfiles}. The morphology
  does not change significantly with energy. This is qualitatively
  compatible with the results of the spatially resolved spectral
  analysis (see section~\ref{subsec:spectra}).

  \subsection{Energy spectra}\label{subsec:spectra}
\begin{figure}
    \resizebox{\hsize}{!}{\includegraphics{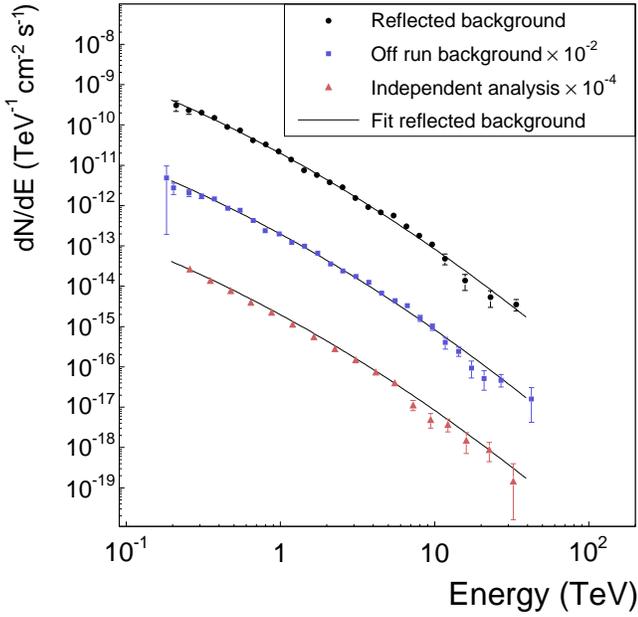}}
    \caption{Shown are three spectra that were produced to explore the
    systematic uncertainties. The alternative spectra (blue squares
    and red triangles) were scaled by factors of $10^{-2}$ and
    $10^{-4}$, respectively, for presentation reasons. \textbf{Upper
    set:} Standard analysis and standard background modelling using a
    background position from the same field of view, with the exact
    same shape and distance to the pointing direction as the signal
    region, but opposite to it, on the other side of the pointing
    direction. \textbf{Middle set:} Alternative spectral analysis
    technique (event-wise effective area weighting, see
    Sect.~\ref{subsec:SpectralAnalysis}) with an independent
    background estimate taken from OFF runs. The background regions in
    the OFF data were again selected such that they have the same
    shape and distance to the pointing direction as the signal region
    in the \rxj\ observations. \textbf{Lower set:} Spectrum produced
    applying an independent analysis chain. The background was
    determined similarly as for the upper set. A third spectral
    analysis technique was applied here, described in
    \citet{Piron}. Plotted as black line on top of all three spectra
    to guide the eye is the best fit of a power law with energy
    dependent photon index to the spectrum shown in the upper set (see
    Table~\ref{table:fits} for details). The error bars on the
    spectral points denote $\pm 1 \sigma$ statistical errors.}
    \label{fig:SpectraCheckI}
  \end{figure}

  The energy spectrum of \rxj\ was measured with \hess\ over a large
  energy range. Systematic tests on the shape and characteristics of
  the energy spectrum of the whole SNR included application of a
  slightly different spectral analysis technique, different background
  models, analysis in distinct data subsets like small and large
  zenith angles and the five observation positions, analysis applying
  different cuts on image intensity and telescope multiplicity,
  investigation of the influence of the exact binning, of the energy
  estimation and the fit to the effective area histograms obtained
  from simulations. Furthermore, the results were cross-checked with
  the independent calibration and analysis scheme mentioned in
  Sect.~\ref{subsec:morphology}. Representative examples for these
  systematic tests are shown in Fig.~\ref{fig:SpectraCheckI}, where the
  spectrum of the whole SNR is plotted using the standard background
  estimation from the same field of view, compared to a completely
  independent background estimation based on OFF runs, and to the
  independent analysis chain~\citep{ModelPaper}. The three spectra are
  found to be well compatible with each other.

  In order to obtain a quantitative estimate for the systematic error
  on each flux point, the background estimation, the spectral analysis
  technique (event-wise effective area weighting or average effective
  area determination per data run) and the absolute energy scale of
  the experiment were considered as dominant contributions. For the
  background modelling, an uncertainty of $\Delta\alpha = 1\%$,
  $\alpha$ being the background normalisation factor, was derived from
  the \rxj\ data set by comparing the standard analysis to the
  analysis using an independent background model derived from OFF
  runs, as mentioned above. It should be noted that the 1\% variation
  found in the total number of background counts is compatible with
  being due to a statistical variation and thus must be regarded as an
  upper limit. Taking the systematic uncertainty due to the background
  level and the two spectral analysis techniques, an energy dependent
  systematic error was obtained by analysing the data set six times --
  scaling $\alpha$ by $(1. \, + \, [-1,0,+1] \times \Delta\alpha)$ and
  applying both analysis techniques. The RMS of the resulting six flux
  points in each energy bin was taken as the systematic
  uncertainty. The uncertainty in the energy scale is a global
  uncertainty which might cause a shift of the whole spectrum to lower
  or larger energies. It is due to uncertainties in the atmospheric
  transmission models used in
  simulations~\citep[see][]{StefanHessTrigger} and uncertainties in
  the light collection efficiencies of the telescopes. The combined
  error is estimated to be $20\%$.

  When fitting a power law with index $\Gamma$ to the spectrum, the
  systematic error on the integral flux obtained from the fit function
  is conservatively estimated to be $25\%$, on the fit index it is
  $\Delta\Gamma = 0.1$.

  \begin{figure}
    \resizebox{\hsize}{!}{\includegraphics{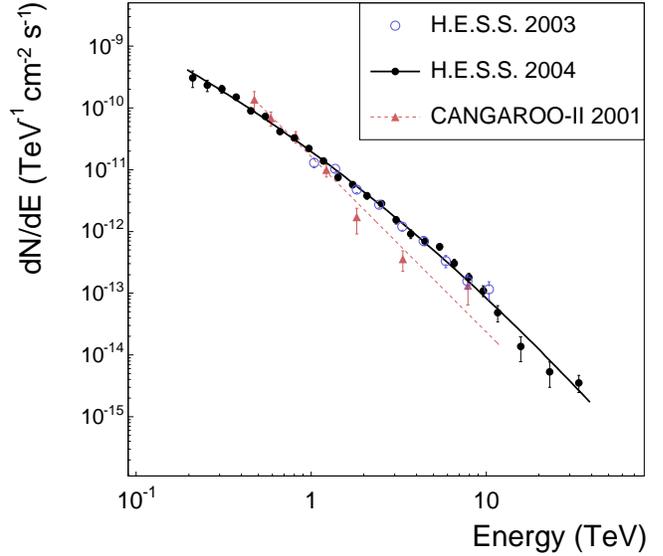}}
    \caption{Differential gamma-ray energy spectrum of \rxj, for the
    whole region of the SNR (solid black circles). The best fit of a
    power law with energy dependent photon index is plotted as black
    line. For comparison the \hess\ 2003 data points are also shown
    (blue open circles). Note the vast increase in energy coverage due
    to the increased sensitivity of the complete telescope array. The
    spectrum ranges now from 190~GeV to 40~TeV, spanning more than two
    decades in energy. The data points reported by the CANGAROO-II
    collaboration~\citep{CANGII} for the northwestern part of the
    remnant are also shown as red triangles, the corresponding best
    fit result as dashed red line. Error bars are $\pm 1 \sigma$
    statistical errors.}
    \label{fig:Spectrum}
  \end{figure}
  The spectrum of the whole SNR was determined by integrating events
  within $0.65\degr$ radius around the centre of the SNR,
  $\alpha_{\mathrm{J2000}} \, = \, \mathrm{17h13m33.6s}$,
  $\delta_{\mathrm{J2000}} \, = \, -39\degr 45\arcmin 36\arcsec$. OFF
  events were selected from a reflected region in the same field of
  view (see Sect.~\ref{subsec:SpectralAnalysis} for explanation and
  Fig.~\ref{fig:OffRegions} for illustration). To ensure optimum match
  in the offset distributions of ON and OFF events, runs taken
  directly on the source, where no appropriate OFF region can be
  selected in the same field of view, were discarded from the spectral
  analysis. Accordingly, the total live time reduced slightly to 30.5
  hours. A size cut of 80 photo-electrons was applied for the spectral
  analysis. This results in a cumulative significance of $31\sigma$
  corresponding to $\approx 15400$ excess events (normalisation factor
  $\alpha \, = \, 1.3$).

  The resulting spectrum of the whole SNR is shown in
  Fig.~\ref{fig:Spectrum}. The data is in excellent agreement with the
  previous measurement in 2003, which covered the energy range from
  1~TeV to 10~TeV. The latest data span more than two orders of
  magnitude in energy, from 190~GeV to 40~TeV. The best fit of a power
  law with energy dependent photon index is plotted (the exact formula
  is given below). It describes the data reasonably
  well. Table~\ref{table:fits} summarises fits of different spectral
  shapes to the data. Three alternative shapes have been used: a power
  law with an exponential cutoff $E_{\mathrm{c}}$,
  \begin{eqnarray*}
    \frac{d\mathrm{N}}{d\mathrm{E}} = I_0\ \left(
    \frac{E}{1~\mathrm{TeV}}\right) ^ {-\Gamma}\ \exp
    \left(-\frac{E}{E_{\mathrm{c}}}\right) \, \mathrm{,}
  \end{eqnarray*}
  a power law with an energy dependent exponent,
  \begin{eqnarray*}
    \frac{d\mathrm{N}}{d\mathrm{E}} = I_0\
    \left(\frac{E}{1~\mathrm{TeV}}\right) ^ {-\Gamma + \, \beta\ 
    \log \frac{E}{1~\mathrm{TeV}}} \, ,
  \end{eqnarray*}
  and a broken power law (transition from $\Gamma_1$ to $\Gamma_2$ at
  break energy $E_{\mathrm{B}}$, $S$ quantifies the sharpness of the
  transition),
  \begin{eqnarray*}
    \frac{d\mathrm{N}}{d\mathrm{E}} = I_0\
      \left(\frac{E}{E_{\mathrm{B}}}\right) ^
      {-\Gamma_1}\ \left( 1 + \left(\frac{E}{E_{\mathrm{B}}}\right) ^ {1 / S } \right) ^
      {\, S \, (\Gamma_2 - \Gamma_1)} \, .
  \end{eqnarray*}
  In all cases, $I_0$ is the differential flux normalisation, the
  energies $E$ are normalised at 1~TeV and photon indices are
  specified with $\Gamma$.

  All three alternative shapes describe the data significantly better
  than the pure power law. However, among the alternative spectral
  shapes, none is significantly favoured over the others. At the
  highest energies, above 10~TeV, there is still a significant
  gamma-ray flux in excess of $6\sigma$. It should be noted, though,
  that in order to draw strong conclusions about the high-energy shape
  of the spectrum, more data with better statistics at the high-energy
  end of the spectrum are needed.
  \begin{table*}
    \caption{Fit results for different spectral models. The
    differential flux normalisation $I_0$ and the integral flux above
    1~TeV $(I(>1~\mathrm{TeV}))$ are given in units of $10^{-12} \,
    \mathrm{cm}^{-2} \, \mathrm{s}^{-1} \, \mathrm{TeV}^{-1}$ and
    $10^{-12} \, \mathrm{cm}^{-2} \, \mathrm{s}^{-1}$,
    respectively. The power-law fit is clearly an inappropriate
    description of the data, a power law with an exponential cutoff
    (row 2), a power law with an energy dependent photon index (row
    3), and a broken power law (row 4; in the formula, the parameter
    $S = 0.4$ describes the sharpness of the transition from
    $\Gamma_1$ to $\Gamma_2$ and it is fixed in the fit) are equally
    likely descriptions of the \hess\ data. Note that when fitting a
    broken power law to the data, some of the fit parameters are
    highly correlated.}
    \label{table:fits}
    \centering
    \begin{tabular}{*{7}{l}}
      \hline
      \hline
      Fit Formula & \multicolumn{4}{c}{Fit Parameters} &
      $\chi^2$~(d.o.f.) & $I(>1~\mathrm{TeV})$ \\
      \hline\\[-0.6ex]
      $I_0\ E ^ {-\Gamma}$ & $I_0 = 17.1
      \pm 0.5$ & $\Gamma = 2.26 \pm 0.02$ & & &
      85.6 (23) & $13.5 \pm 0.4$ \\[1.4ex]
      $I_0\ E ^ {-\Gamma}\ \exp (-E / E_{\mathrm{c}})$ & $I_0 = 20.4
      \pm 0.8$ & $\Gamma = 1.98 \pm 0.05$ & $E_{\mathrm{c}} = 12 \pm
      2 $ & & 27.4 (22) & $15.5 \pm 1.1$ \\[1.4ex]
      $I_0\ E ^ {-\Gamma + \, \beta\ \log E}$ & $I_0 = 19.7
      \pm 0.6$ & $\Gamma = 2.08 \pm 0.04$ & $\beta = -0.30 \pm
      0.04 $ & & 25.5 (22) & $15.6 \pm 0.7$ \\[1.4ex]
      $I_0\ E / E_{\mathrm{B}} ^ {-\Gamma_1}\ \left( 1 + E /
      E_{\mathrm{B}} ^ {1 / S } \right) ^ {\, S \, (\Gamma_2
      - \Gamma_1)}$ & $I_0 = 0.4^{+0.6}_{-0.2}$ & $\Gamma_1
      = 2.06 \pm 0.05$ & $\Gamma_2 = 3.3 \pm 0.5$ & $ E_{\mathrm{B}} =
      6.7 \pm 2.7 $ & 26.2 (21) & $15.4 \pm 0.8$ \\
    \end{tabular}
  \end{table*}

  The spectrum reported by the CANGAROO-II
  collaboration~\citep{CANGII} for the northwest part of the SNR is
  also shown in Fig.~\ref{fig:Spectrum}. From a power-law fit to the
  data they quoted a photon index $\Gamma = 2.84 \pm 0.15 \,
  (\mathrm{statistical}) \, \pm 0.20 \, (\mathrm{systematic})$ and a
  differential flux normalisation at 1~TeV $I_0 = (1.63 \pm 0.15 \pm
  0.32) \times 10^{-11} \, \mathrm{cm}^{-2} \, \mathrm{s}^{-1} \,
  \mathrm{TeV}^{-1}$. The difference between the two spectra is
  somewhat larger than the quoted errors of the measurements. However,
  the CANGAROO-II spectrum is only for a part of the remnant. Moreover
  we note that the CANGAROO-II collaboration has recently revised
  their systematic errors upwards. For example, the Galactic Centre
  photon index, which was initially given as $4.6 \pm
  0.5$~\citep{CangGalCen}, was recently quoted as
  $4.6_{-1.2}^{+5.0}$~\citep{KatagiriVelaJr}.

  \begin{figure}
    \resizebox{\hsize}{!}{\includegraphics{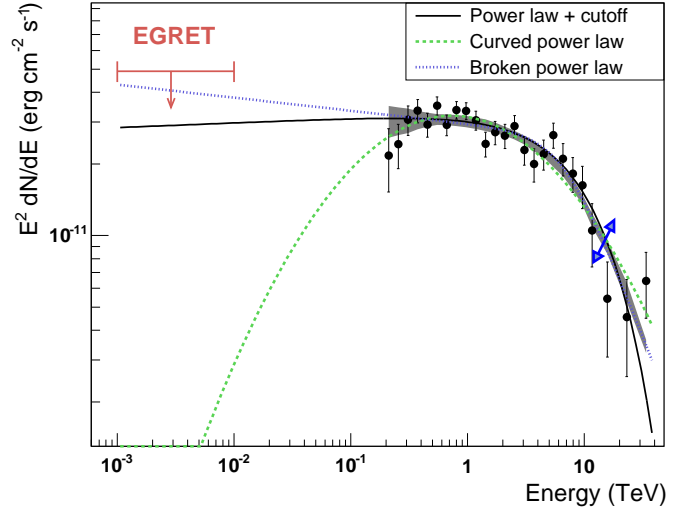}}
    \caption{\hess\ energy spectrum of \rxj. Plotted are the \hess\
    points with their $\pm 1 \sigma$ statistical errors in an energy
    flux diagram. The three curves (specified in
    Table~\ref{table:fits}) are the best fit results of a power law
    with an exponential cutoff, a power law with energy dependent
    photon index, and a broken power law, extrapolated to 1~GeV to
    enable comparisons with the EGRET upper limit in the range of
    1--10~GeV. The shaded grey band represents the systematic
    uncertainty on the measurement, originating from the uncertainty
    on the background estimation. The blue arrow indicates the 20\%
    systematic uncertainty on the energy scale, which might shift the
    whole curve in the given direction.}
    \label{fig:SpectrumSysError}
  \end{figure}
  Figure~\ref{fig:SpectrumSysError} illustrates the three spectral
  shapes that were found to describe the data reasonably well. The
  three curves are extrapolated to 1~GeV to compare them with the
  EGRET upper limit on the energy flux of $4.9 \times
  10^{-11}~\mathrm{erg}~\mathrm{cm}^{-2}~\mathrm{s^{-1}}$, ranging
  from 1~GeV to 10~GeV, centred at 2.9~GeV. The limit was determined
  at the \hess\ position of \rxj\ by modelling and subtracting the
  known EGRET source \object{3EG~1714$-$3857}~\citep{Hartmann},
  assuming that \object{3EG~1714$-$3857} is not linked to the
  gamma-ray emission of \rxj. Since the \hess\ location is in close
  vicinity of \object{3EG~1714$-$3857} (actually it is overlapping),
  this procedure could only be carried out successfully above 1
  GeV. The systematic error band for the \hess\ data was obtained as
  described above. It is centred on the mean value of the three fit
  curves and represents the systematic error due to background
  uncertainties only. The energy scale is an energy independent
  uncertainty; its scale and the direction the curve is shifted to are
  marked with a blue arrow at one representative position (at
  15~TeV). It is worth noting that the systematic uncertainty on the
  background has a considerable impact on the first few flux points
  because of the smaller signal-to-noise ratio (as compared to points
  at higher energies). For the spectrum shown here the systematic
  uncertainty is $\approx 18\%$ for the two lowest-energy points; it
  decreases rapidly with increasing energy being well below $10\%$ at
  350~GeV.
  
  \begin{figure*}
    \centering
    \includegraphics[draft=false,width=17cm]{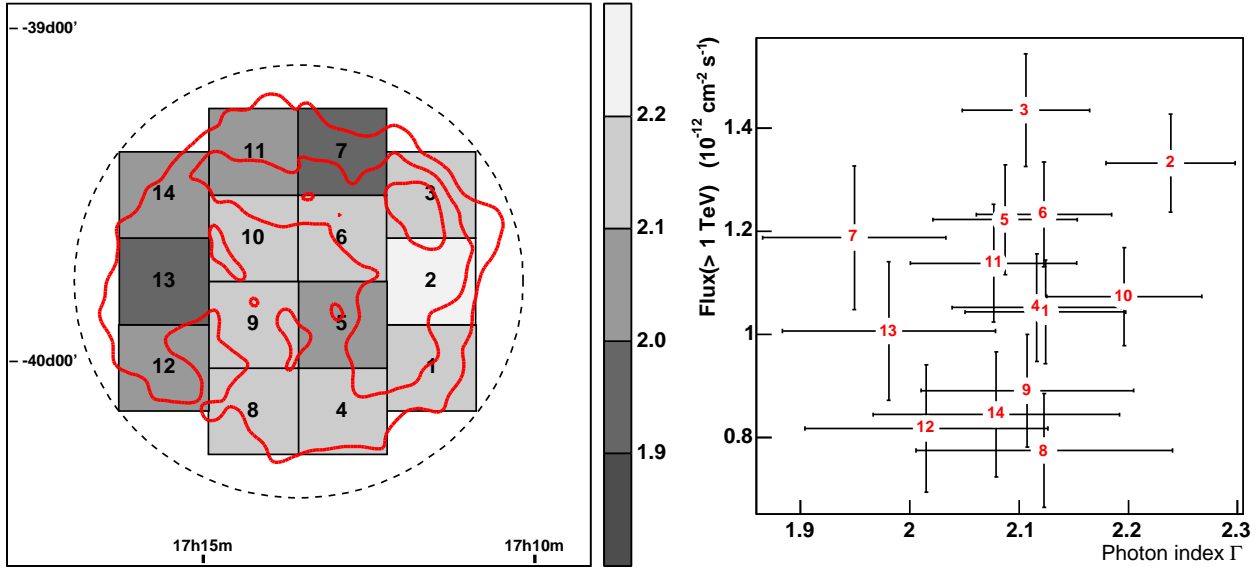}
    \caption{The image illustrates the results of the spatially
      resolved spectral analysis. \textbf{Left part:} Shown in red are
      gamma-ray excess contours from Fig.~\ref{fig:Size200Image},
      linearly spaced at 30, 60, and 90 counts. Superimposed are the
      14 boxes (each $0.26\degr \times 0.26\degr$ in dimension) for
      which spectra were obtained independently. The dashed line is
      the $0.65\degr$ radius circle that was used to integrate events
      to produce a spectrum of the whole SNR. The photon index
      obtained from a power-law fit in each region is colour coded in
      bins of 0.1. The ranges of the fits to the spectra have been
      restricted to a maximum of 8~TeV (see Table~\ref{table:regions}
      ). \textbf{Right part:} Plotted is the integral flux above 1~TeV
      against the photon index, for the 14 regions the SNR was
      sub-divided in. The error bars are $\pm 1 \sigma$ statistical
      errors. Note that systematic errors of 25\% on the flux and 0.1
      on the photon index are to be assigned to each data point
      additionally.}
    \label{fig:IndicesRegions}
  \end{figure*}
  The results of the spatially resolved spectral analysis are shown in
  Fig.~\ref{fig:IndicesRegions}. It shows the distribution of photon
  indices over the SNR resulting from a power-law fit. The spectra
  were determined in rectangular regions, denoted 1--14, each
  $0.26\degr \times 0.26\degr$ in dimension. The fit range was limited
  to 8~TeV to account for (and avoid when fitting) the deviation from
  a power law seen in the spectrum of the whole
  remnant. Table~\ref{table:regions} summarises the fit results. There
  is a significant flux variation over the SNR. From the brightest
  region in the northwest to a relatively dim one in the central part,
  the flux varies by more than a factor of two. There is no
  significant difference in spectral shape apparent, the photon
  indices agree with each other within statistical and systematic
  errors. The distribution of photon indices has a mean value of 2.09
  with a root-mean-square of 0.07. This is well compatible with the
  spectrum of the whole SNR when the fit range is also restricted to
  maximum 8~TeV for consistency (first row in
  Table~\ref{table:regions}). If one adds up the integral fluxes above
  1~TeV of the individual regions, a flux of $15.1 \times 10^{-12} \,
  \mathrm{cm}^{-2} \, \mathrm{s}^{-1}$ is obtained, 5\% less than the
  flux of the whole SNR (with the restriction of the fit range). This
  is in excellent agreement with expectations; the boxes as they are
  plotted in Fig.~\ref{fig:IndicesRegions} cover the region of \rxj\
  with significant gamma-ray excess almost completely.
  \begin{table}
    \caption{Fit results for distinct regions of the SNR. Given are
      for each region the photon index resulting from a power-law fit,
      the best-fit $\chi^2$ and the number of degrees of freedom
      (d.o.f.), the integral flux above 1~TeV and the significance of
      the excess events in units of standard deviation $(\sigma)$. The
      background for each region was determined from the same field of
      view, as described in Sect.~\ref{subsec:SpectralAnalysis}, for
      each region separately, and hence the background estimates for
      different regions are not independent. The first row is the fit
      result of the whole SNR for comparison. For the whole table, the
      upper fit range was restricted to 8~TeV to avoid biases due to
      the deviation from a power law at high energies.}
    \label{table:regions}
    \centering
    \begin{tabular}{*{5}{c}}
      \hline
      \hline
      Region & $\Gamma$ & $\chi^2$~(d.o.f.) &  $I(>1~\mathrm{TeV})$ & Excess\\
      & & & $(10^{-12} \, \mathrm{cm}^{-2}~\mathrm{s}^{-1})$ & $(\sigma)$\\
      \hline
      All & $2.12 \pm 0.03$ & 24.5 (18) & $15.9 \pm 0.6$ & 30.8\\
      1 & $2.12 \pm 0.07$ & 34.7 (18) & $1.05 \pm 0.13$ & 12.9\\
      2 & $2.24 \pm 0.06$ & 26.0 (17) & $1.34 \pm 0.10$ & 17.2\\
      3 & $2.11 \pm 0.06$ & 30.2 (18) & $1.45 \pm 0.13$ & 16.7\\
      4 & $2.10 \pm 0.08$ & 15.7 (18) & $1.06 \pm 0.12$ & 11.5\\
      5 & $2.09 \pm 0.07$ & 12.6 (18) & $1.22 \pm 0.11$ & 13.3\\
      6 & $2.13 \pm 0.06$ & 35.7 (17) & $1.23 \pm 0.12$ & 14.1\\
      7 & $1.95 \pm 0.08$ & 9.4 (16) & $1.19 \pm 0.12$ & 10.9\\
      8 & $2.11 \pm 0.12$ & 13.8 (14) & $0.78 \pm 0.11$ & 8.0\\
      9 & $2.11 \pm 0.10$ & 12.5 (16) & $0.89 \pm 0.11$ & 8.7\\
      10 & $2.19 \pm 0.07$ & 24.8 (17) & $1.09 \pm 0.10$ & 14.1\\
      11 & $2.08 \pm 0.08$ & 11.6 (15) & $1.13 \pm 0.11$ & 11.8\\
      12 & $2.01 \pm 0.11$ & 8.4 (16) & $0.81 \pm 0.12$ & 8.2\\
      13 & $1.98 \pm 0.10$ & 10.7 (15) & $1.00 \pm 0.14$ & 9.8\\
      14 & $2.08 \pm 0.11$ & 9.9 (15) & $0.84 \pm 0.12$ & 9.4\\
      \hline
    \end{tabular}
  \end{table}
  
  As can be seen from Fig.~\ref{fig:IndicesRegions}, right part,
  there is no correlation of the gamma-ray flux and the photon index
  visible in the data. This, together with the absence of any change
  in the spectral shape, is a remarkable difference between the
  gamma-ray and X-ray data. The spectral variation in X-rays was found
  to be much larger~\citep[see][]{CassamXMM}.
  
  \section{\rxj\ at other wavelengths}\label{sec:Multiwavelength}
  \subsection{Comparison with X-rays}\label{subsec:XrayData}
  \begin{figure}
    \resizebox{\hsize}{!}{\includegraphics[draft=false]{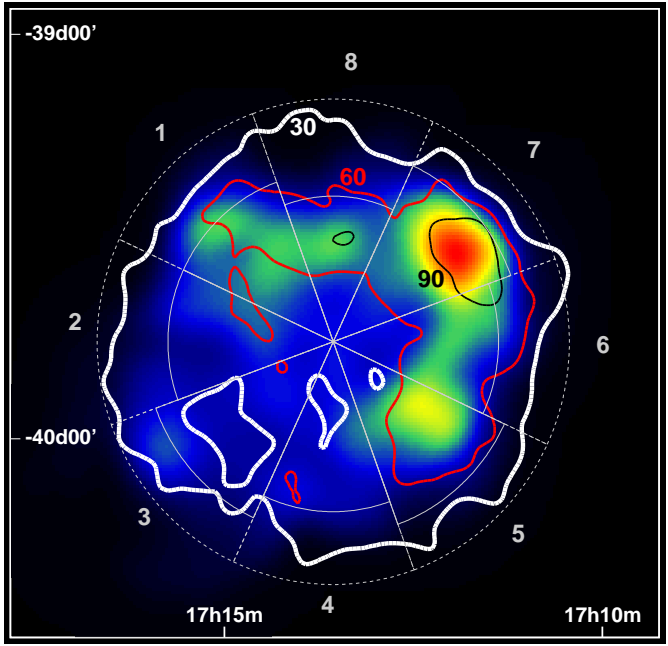}}
    \caption{ASCA X-ray~\citep[1--5~keV band,][]{UchiyamaPrivate} image
    of \rxj, overlaid with contours of the smoothed,
    acceptance-corrected \hess\ gamma-ray image. The coloured contour
    levels are labelled and linearly spaced at 30, 60, and 90
    counts. Drawn as gray thin lines are eight wedge-shaped regions
    for which the radial profiles are compared to each other in
    Fig.~\ref{fig:RadialProfiles}. Note that in the ASCA image, most
    of the regions (faint solid lines) do not reach as far as in the
    \hess\ image (faint dashed lines), accounting for the limited
    field of view of ASCA, whose coverage did not always extend to the
    boundaries of the SNR. As explained in the main text, the ASCA
    image was smoothed to match the \hess\ point-spread function to
    enable comparison of the two images.}
    \label{fig:HessAsca}
  \end{figure}
  \begin{figure*}
    \centering
    \includegraphics[draft=false,width=17cm]{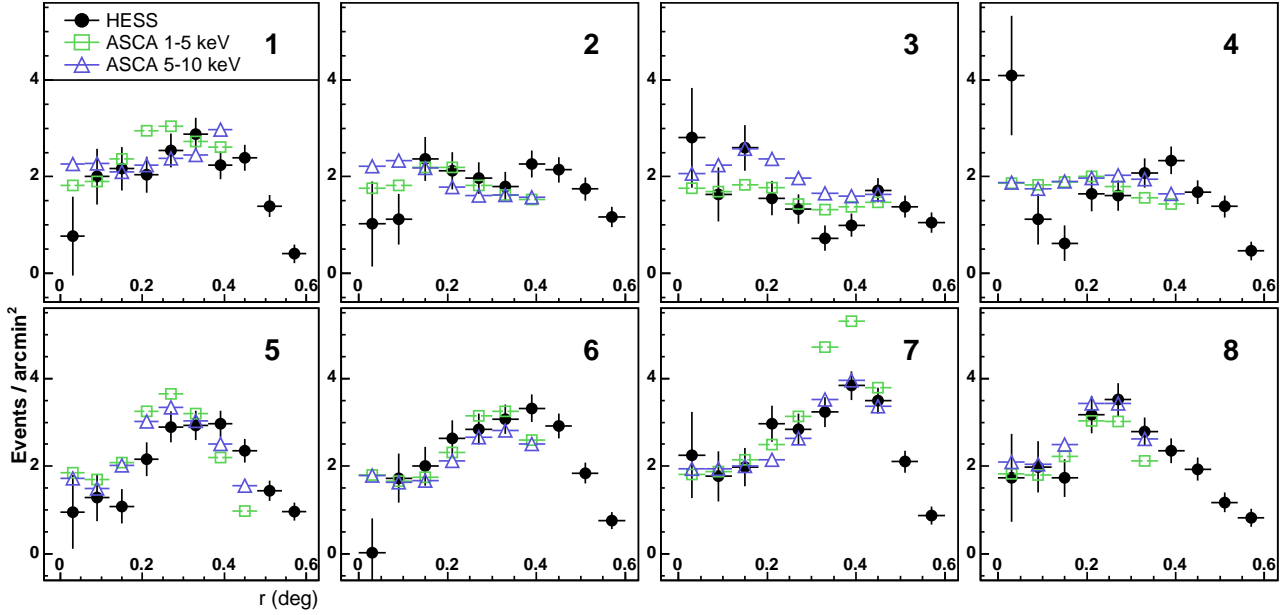}
    \caption{Radial profiles for the eight regions marked in
    Fig.~\ref{fig:HessAsca}. Plotted are \hess\ excess counts per unit
    solid angle (solid circles) as a function of distance $r$ in
    degrees to the centre of the SNR, compared to soft (1--5~keV) and
    hard (5--10~keV) X-ray data. All data were corrected for relative
    acceptance. There is a very good general agreement between the keV
    and TeV data sets. The most pronounced differences appear in
    regions 4, where the TeV flux drops almost to zero at $\approx
    0.15\degr$ from the centre, and 7, where a pronounced peak appears
    in the 1--5~keV X-ray data, which repeats neither in the 5--10~keV
    X-ray nor the TeV data. Note that the X-ray emission includes
    diffuse Galactic X-ray emission.}
    \label{fig:RadialProfiles}
  \end{figure*}
  There is a striking similarity between the X-ray and the gamma-ray
  image of \rxj, as they are shown in Fig.~\ref{fig:HessAsca} for the
  1-5~keV X-ray band. The overall morphology appears to be very
  similar, the brightest spots in both images are distributed on the
  shell, especially in the west. For a detailed comparison one must
  take into account the slightly better resolution of ASCA compared to
  \hess. For that purpose, the ASCA image was smoothed beyond the
  point-spread function of the instrument in order to match the \hess\
  resolution. The smoothing radius was determined empirically by
  smoothing the ASCA point-spread function and comparing it with the
  \hess\ point-spread function. An optimum match was obtained for a
  smoothing radius of $0.037\degr$. Furthermore, both images in
  Fig.~\ref{fig:HessAsca} are corrected for relative detector
  acceptance. The X-ray-bright central point source \object{1WGA
  J1713.4-3949}, which was argued to be the compact relic of the
  supernova progenitor~\citep[see for example][]{CassamXMM}, was
  removed from the X-ray image for the purpose of comparison with
  gamma rays.

  After degradation of ASCA's resolution, the data were compared to
  each other in eight wedge-shaped regions as they are drawn in
  Fig.~\ref{fig:HessAsca}. In each wedge, radial profiles, that is,
  the number of counts per unit solid angle as function of distance to
  the centre, were determined. To account for the differences in the
  absolute count level the X-ray images were scaled by a normalisation
  factor, which has been calculated as the ratio of TeV counts,
  integrated in a rectangle encompassing the SNR (and within the ASCA
  field of view), to keV counts, integrated in the same rectangle. The
  result is shown in Fig.~\ref{fig:RadialProfiles} for all eight
  wedges. The overall good agreement in shape of the profiles is
  clearly visible, differences appear only at a few places, for
  example in region 4 and 7. For a quantitative statement on the
  compatibility of the two data sets, however, one would have to model
  and subtract the contribution from Galactic diffuse X-ray emission
  in the ASCA image, which amounts presumably to $10\%$ or less in the
  X-ray bright parts of the SNR, but might increase to $\approx 30\%$
  in the faint parts in the east.

  The interesting question of the boundaries of the SNR and if they
  are the same in X-rays and gamma rays can unfortunately not be
  addressed with the ASCA data set due to limited sky coverage.

  \subsection{CO and radio observations}\label{subsec:CORadioData}
  \begin{figure*}
    \centering
    \includegraphics[draft=false,width=17cm]{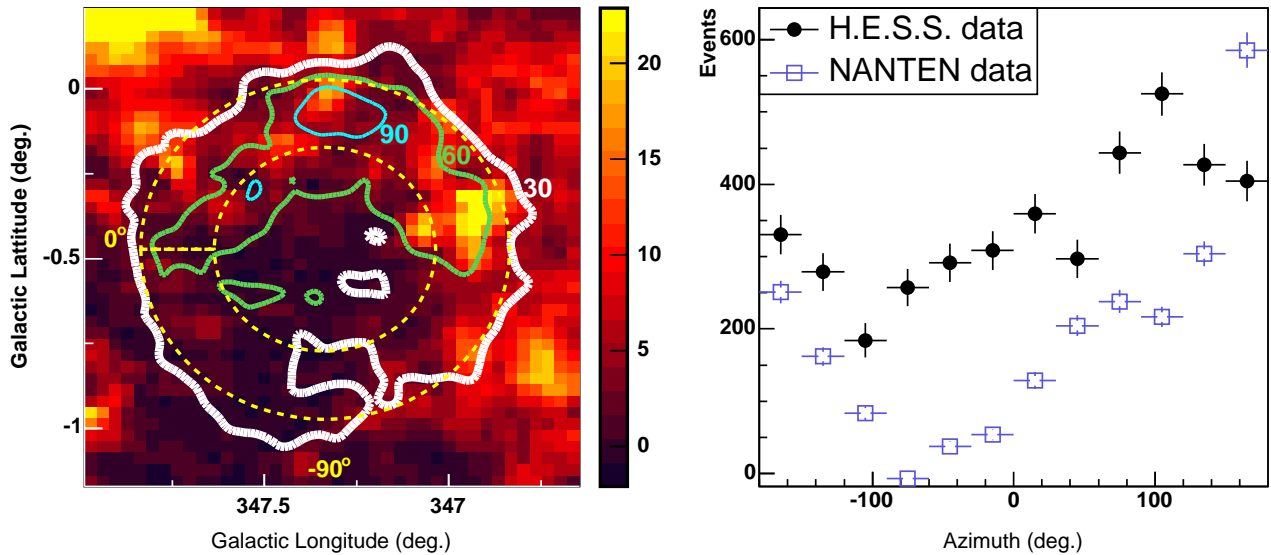}
    \caption{\textbf{Left panel:} Shown are the intensity distribution
    of CO $(J = 1-0)$ emission~\citep{FukuiUchiyama} (linear colour
    scale in units of $\mathrm{K}~\mathrm{km}~\mathrm{s}^{-1}$,
    truncated at a value of 23 to highlight important features),
    derived by integrating the CO spectra in the velocity range from
    $-11~\mathrm{km}~\mathrm{s}^{-1}$ to
    $-3~\mathrm{km}~\mathrm{s}^{-1}$ (which corresponds to
    $0.4~\mathrm{kpc}$ to $1.5~\mathrm{kpc}$ in space). Overlaid are
    coloured contours of the \hess\ gamma-ray excess image. The levels
    are labelled and linearly spaced at 30, 60, and 90 counts. Note
    that the image is shown in Galactic coordinates. \textbf{Right
    panel:} Azimuth profile plot, that is, number of counts as a
    function of the azimuthal angle, integrated in a $0.2\degr$-wide
    ring covering the shell of \rxj\ (dashed yellow circle in the
    left-hand panel). Plotted are the \hess\ gamma-ray and the NANTEN
    data set.}
    \label{fig:COImage}
  \end{figure*}
  CO data at 2.6~mm wavelength of \rxj\ and its surroundings were
  taken with the 4-m, mm and sub-mm telescope NANTEN in
  2003~\citep{FukuiUchiyama}. Based on positional coincidences of CO
  and X-ray peaks and (in velocity space) shifted CO peaks,
  \citet{FukuiUchiyama} concluded that the SNR blast wave is
  interacting with molecular clouds situated on its western side at a
  distance of $\approx 1~\mathrm{kpc}$. Further possible support for
  this scenario was recently published in \citet{Moriguchi}, where
  high gas excitations are reported for this part, which could arise
  from heating of the molecular gas by the shock wave. The CO
  intensity distribution in the corresponding velocity interval is
  shown in Fig.~\ref{fig:COImage}, together with \hess\ gamma-ray
  excess contours. One notes that in the central and southeastern part
  of the SNR the CO emission becomes very faint or is completely
  absent.  Apart from that, there are local CO maxima that coincide
  with TeV-bright parts on the western side of the SNR. The azimuth
  profile plotted on the right-hand side of Fig.~\ref{fig:COImage}
  illustrates a global agreement between the two measurements, regions
  with low gamma-ray flux reveal also low CO intensity, but there is
  no exact proportionality between the two measurements for the shell
  region of \rxj. Taking the peak values, one notes that they are
  shifted with respect to each other and that the gamma-ray flux
  varies by a factor of about three, whereas the CO intensity drops by
  roughly a factor of 100 in the central-southeastern part.
  
  \begin{figure}
    \resizebox{\hsize}{!}{\includegraphics[draft=false]{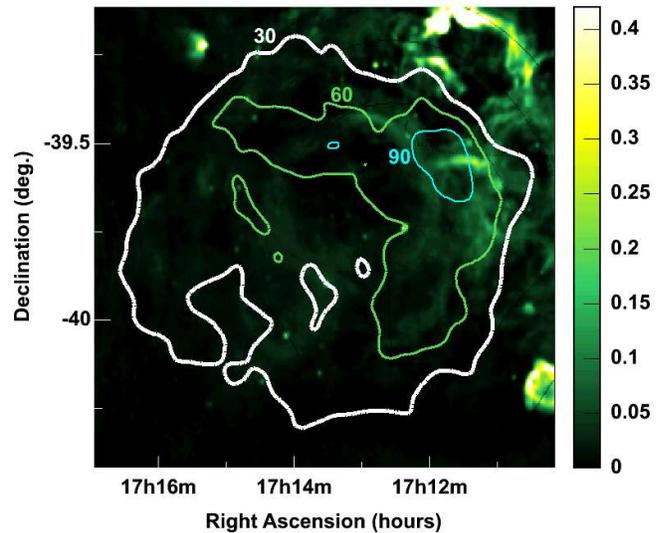}}
    \caption{1.4~GHz ATCA radio image~\citep[][ courtesy of
    P.Slane]{Lazendic}. The linear colour scale is in units of
    $\mathrm{Jy}~\mathrm{beam}^{-1}$. Overlaid are coloured contour
    lines of the \hess\ gamma-ray excess image.}
    \label{fig:RadioImage}
  \end{figure}
  Figure~\ref{fig:RadioImage} shows a comparison of the 1.4~GHz radio
  image obtained with ATCA~\citep{Lazendic} and the \hess\ gamma-ray
  excess contours. The SNR is very faint in the radio band, there are
  two faint arc-like structures of emission to the west of \rxj,
  almost perpendicular to each other, one of them directly coincident
  with the brightest TeV emission region. There is no notable
  resemblance between the two wavelength regimes. Spectral analysis of
  the X-ray- and TeV-bright northwestern part of the SNR shell yields
  a spectral index of $0.50 \pm 0.40$, derived from two flux values
  taken at 1.4~GHz and 2.5~GHz. This measurement is used further down
  when comparing the \hess\ spectral data to broadband models.

  \section{Possible emission processes}\label{sec:Modelling}
  One of the key issues in the interpretation of the observed
  gamma-ray emission is the identification of the particle population
  responsible for the generation of the gamma rays. The close
  correlation between X-rays and gamma rays might indicate an
  electronic origin; models of supernova remnants as Galactic
  cosmic-ray sources, on the other hand, suggest that primarily a
  hadronic component from pion decays exists. To identify the
  different contributions, the wide-band electromagnetic spectra from
  radio to multi-TeV gamma-rays must be compared to model
  calculations.

  In the literature, different schemes are employed to model broadband
  emission from SNRs. Phenomenologically oriented
  models~\citep{Mastichiadis96,Aharonian99} start by ad hoc assuming
  particle acceleration spectra - usually as power laws with a cutoff
  - to derive particle spectra taking into account energy losses and
  then calculate the electromagnetic spectrum with additional
  assumptions concerning the local magnetic field, the radiation
  fields which serve as target for the IC process, and the gas
  density. Spectral parameters are either taken from acceleration
  models, resulting in a spectral index around 2, or determined from
  data. More sophisticated gamma-ray models account for the non-linear
  effects arising from the interaction of the accelerated particles
  with the shocked supernova shell, which result in deviations from
  pure power laws, with spectra flattening at higher
  energies~\citep{Berezhko97,Baring99}.

  The original discovery paper of VHE gamma-ray emission from
  \rxj~\citep{CANGI} claimed electrons as the likely source particle
  population. However, it soon became evident that a consistent
  modelling of the spectra is hard to achieve in simple one-zone
  models. Apart from the choice of the electron spectrum, the only
  free parameter is the magnetic field strength, which controls the
  spacing of the synchrotron and IC peaks in the SED together with
  their relative intensities and -- one should add -- the amount of
  radiative cooling of the accelerated component. \citet{CANGII} noted
  that for modest magnetic fields -- $B$ equal to a few $\mu$G -- the
  measured intensity ratios are reproduced but the gamma-ray spectra
  are much too hard. Using higher fields, one can match the gamma-ray
  spectra at the expense of dramatically increased X-ray yields. While
  the \hess\ data differ from the CANGAROO-II data both in terms of
  the region covered and the exact values for flux and index, this
  conclusion for the electronic scenario remains basically valid. The
  agreement can be improved by introducing an additional parameter to
  decouple the X-ray intensity and the spectral shape, namely the
  magnetic field filling factor which allows the X-ray flux to be
  tuned without change of the spectra. With very small filling factors
  of 0.001~\citep{Pannuti} to 0.01~\citep{Lazendic}, difficult to
  justify physically, the X-ray and the CANGAROO-II gamma-ray spectra
  can be described for magnetic fields around 10~$\mu$G to 15~$\mu$G
  in the emitting region. This latter approach is not followed here.

  The validity of electronic models could be judged more easily if the
  magnetic field values in the remnant were known. For typical shock
  compression ratios around 4 and pre-shock interstellar fields of a
  few $\mu$G, fields of 10~$\mu$G to 15~$\mu$G are at the lower limit
  of the expected range; mechanisms of dynamical field amplification
  in non-linear shocks~\citep{LucekBell,BellLucek,Bell2004} will
  generally result in higher fields. The narrow filaments visible in
  many high-resolution X-ray images of SNRs~\citep[see,
  e.g.,][]{Bamba2005} have been pointed out to provide means to probe
  magnetic fields~\citep{Vink2003,Berezhko2003}: only relatively high
  fields can result in sufficiently rapid cooling of electrons to make
  such filamentary features possible and visible. On the basis of the
  structures seen in Chandra images in the northwest of
  \rxj~\citep{UchiyamaChandra}, \citet{Voelk2005} have argued that
  fields between 58~$\mu$G and a few 100~$\mu$G might be possible,
  depending on the detailed assumptions about the remnant's
  morphology\footnote{We note that in \citet{HiragaXMM}, Fig.~2, a
  radial profile from the XMM image is shown. It reveals another very
  thin filament-like structure in the west of \rxj\ which is a sign of
  high magnetic field values.}. Such high fields -- likely to be
  present throughout the remnant -- would rule out a leptonic origin
  of VHE gamma rays right away.

  On the basis of the difficulty of accommodating broadband spectra in
  a single-zone electronic model, \citet{CANGII} proposed \rxj\ as the
  first well-identified proton accelerator. This interpretation was
  criticised by \citet{Butt} and \citet{ReimerPohl} since the
  CANGAROO-II spectra, extrapolated to lower energies, would violate
  the flux level of the nearby EGRET source
  \object{3EG~1714$-$3857}~\citep{Hartmann}, which, if not associated
  with \rxj, must then be considered as upper limit on the GeV
  emission. However, the EGRET limit can be circumvented by reducing
  the amount of low-energy protons compared to the $E^{-2}$
  spectrum. This can be achieved by the ad hoc assumption of a
  spectral break, or -- for the CANGAROO-II data -- by assuming a
  flatter overall spectrum with a photon index smaller than 2.

  A very detailed modelling is beyond the scope of this paper; the
  models presented in the following serve mainly to illustrate that
  spectra and energetics can be reproduced with plausible input
  parameters. 

  \subsection{Electronic scenario}
  \begin{figure}
    \resizebox{\hsize}{!}{\includegraphics[draft=false]{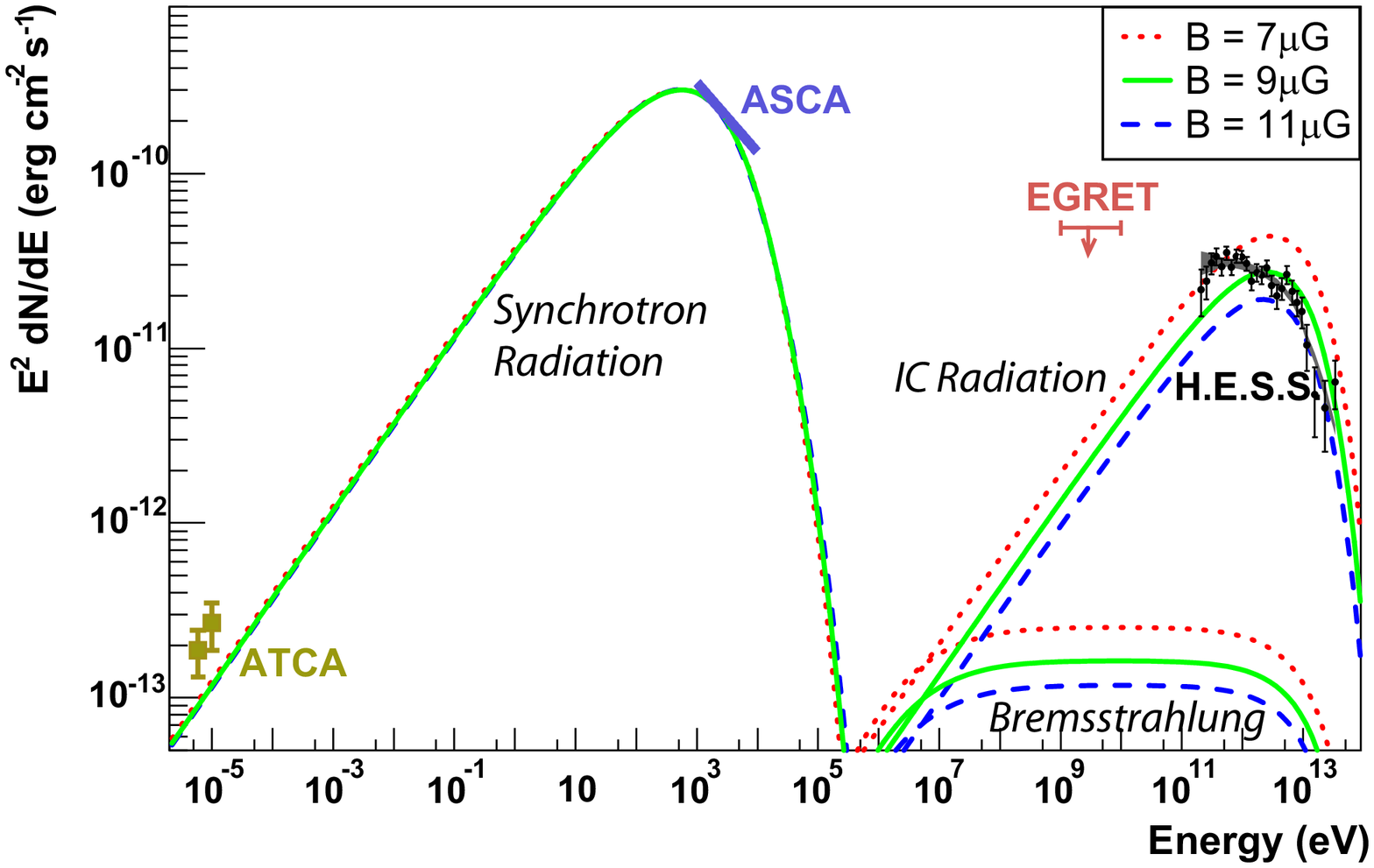}}
    \caption{Broadband SED of \rxj. The ATCA radio data and ASCA X-ray
    data~\citep{HiragaPrivate} for the whole SNR are indicated, along
    with the \hess\ measurement and the EGRET upper limit. Note that
    the radio flux was determined in~\citet{Lazendic} for the
    northwest part of the shell only and was scaled up by a factor of
    two here to account for the whole SNR. The synchrotron and IC
    spectra were modelled assuming a source distance of 1~kpc, an age
    $T$ of 1000~years, a density $n$ of $1~\mathrm{cm}^{-3}$, and a
    production rate of relativistic electrons by the acceleration
    mechanism in the form of a power law of index $\alpha = 2$ and an
    exponential cutoff of $E_0 = 100~\mathrm{TeV}$. Shown are three
    curves for three values of the mean magnetic field:
    $7~\mu\mathrm{G}$, $9~\mu\mathrm{G}$, and $11~\mu\mathrm{G}$, to
    demonstrate the required range of the B field strength for this
    scenario. The electron luminosity is adopted such that the
    observed X-ray flux level is well matched. For the three magnetic
    field values the luminosity $L_{\mathrm{e}}$ is $L_{\mathrm{e}} =
    1.77 \times 10^{37}~\mathrm{erg}~\mathrm{s}^{-1}~
    (7~\mu\mathrm{G})$, $L_{\mathrm{e}} = 1.14 \times
    10^{37}~\mathrm{erg}~\mathrm{s}^{-1}~ (9~\mu\mathrm{G})$, and
    $L_{\mathrm{e}} = 0.81 \times
    10^{37}~\mathrm{erg}~\mathrm{s}^{-1}~ (11~\mu\mathrm{G})$.}
    \label{fig:ElCase1}
  \end{figure}
  In Fig.~\ref{fig:ElCase1} the synchrotron and IC emission from
  relativistic electrons are modelled within the framework of a
  one-zone model in which the electron acceleration and gamma-ray
  emission take place in the same region. It is assumed that the
  primary electrons follow a power law with index $\alpha = 2$ and
  with an exponential cutoff $E_0$,
  \begin{eqnarray*}
    Q(E) = Q_0 E^{-\alpha} \exp(-E/E_0) \, ,
  \end{eqnarray*}
  and that they are produced continuously over a fixed time interval
  $T$ inside a region with given homogeneous distributions of magnetic
  field strength $B$ and ambient gas density $n$. The energy
  distribution of the electrons is then calculated taking into account
  energy losses due to IC and synchrotron emission, Bremsstrahlung and
  ionisation as well as losses due to Bohm diffusion. The broadband
  energy distribution of the source is calculated for an age of $T =
  1000$~years, an average gas density of $n = 1~\mathrm{cm}^{-3}$, and
  a distance to the source of $D = 1~{\mathrm{kpc}}$. For the IC
  emission, canonical interstellar values for the seed photon
  densities were considered: $W_{\mathrm{CMB}} =
  0.25~\mathrm{eV}~\mathrm{cm}^{-3}$ for the cosmic microwave
  background (CMB), $W_{\mathrm{SL}} =
  0.5~\mathrm{eV}~\mathrm{cm}^{-3}$ for optical star light and
  $W_{\mathrm{IR}} = 0.05~\mathrm{eV}~\mathrm{cm}^{-3}$ for infrared
  background light. The absolute electron production rate, $Q_0$, is
  determined from the constraint of matching the observed X-ray flux
  level. Figure~\ref{fig:ElCase1} shows the resulting model curves,
  together with measurements in various wavelength regimes, for three
  different average magnetic field values. From the absolute levels it
  is evident that a magnetic field around $10~\mu \mathrm{G}$ is
  required in order to explain both the X-ray and gamma-ray flux
  levels. On the other hand one notes that such a model with the above
  mentioned parameters does not provide a reasonable description of
  the \hess\ data. The IC peak appears too narrow to reproduce the
  flat TeV emission. The detailed inclusion of non-linear acceleration
  effects should not change the situation very much. They are expected
  to steepen the synchrotron SED above the radio range. Synchrotron
  cooling of the accelerated electrons then tends to produce a
  flat-topped synchrotron and accordingly IC maximum. It is, however,
  a long way to flatten the IC spectrum so extensively at low energies
  as to achieve agreement with the \hess\ spectrum.

  Obviously, the simple model presented here served basically to
  underline the main arguments. Nevertheless, the conclusion that a
  power-law production spectrum fails to simultaneously account for
  the radio, X-ray and gamma-ray data appears to be a generic and
  stable feature; additional parameters are required to decouple
  either the TeV and X-ray/radio fluxes - such as a $filling factor$ -
  or the X-ray and radio spectra - such as an ad-hoc spectral break,
  which for the given source age and magnetic field can not be
  justified as an effect of radiative cooling.

  \subsection{Hadronic scenario}
\begin{figure}
    \resizebox{\hsize}{!}{\includegraphics[draft=false]{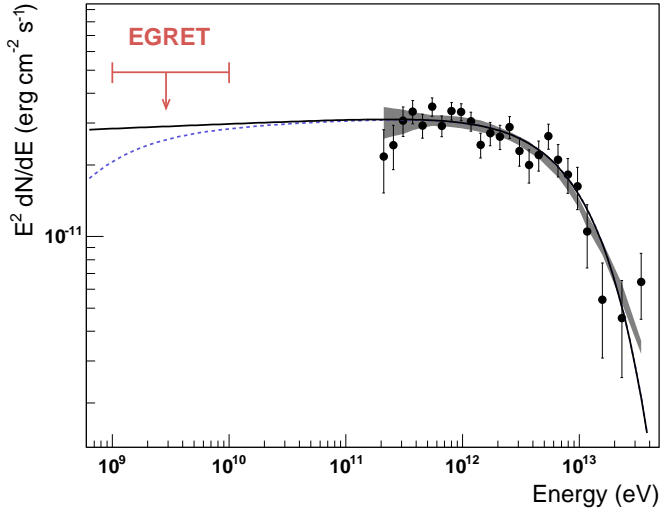}}
    \caption{\hess\ data points plotted in an energy flux
      diagram. They shaded grey band is the systematic error band for
      this measurement (see Sect.~\ref{subsec:spectra}). The black
      curve is the best fit of a power law with exponential cutoff to
      the data, extrapolated to lower energies. The dashed blue curves
      is the same function, but it takes the $\pi^0$ kinematics into
      account. The EGRET upper limit from 1~GeV to 10~GeV is plotted
      as red arrow.}
    \label{fig:PlotHadrons}
  \end{figure}

  Assuming alternatively that nuclear cosmic-ray particles,
  accelerated at the SNR shock, dominantly produce VHE gamma rays,
  theoretically the most plausible differential energy spectrum of
  accelerated nuclei is a concave $E^{-\Gamma (E)}$-type spectrum, due
  to nonlinear back coupling, with a cutoff at gamma-ray energy
  $E_\mathrm{c}$, where $\Gamma$ is decreasing towards higher energies
  (just below the TeV energy range) to a value between 1.5 and 2,
  before the spectrum is steepening again in the cutoff region. In the
  test-particle approximation one expects $\Gamma \simeq 2$. The
  \hess\ spectrum is indeed compatible with such a
  scenario. Figure~\ref{fig:PlotHadrons} shows a $\nu F_\nu$
  representation of the \hess\ data, together with the best-fit curve
  of a power law with an exponential cutoff (see
  Sect.~\ref{subsec:spectra}, Table~\ref{table:fits}), extrapolated to
  small energies. Compared to that a curve is plotted which takes the
  kinematics of the production process of gamma rays, $pp \rightarrow
  \pi^0 \rightarrow \gamma\gamma$, into account. The power law
  spectrum continues to smaller energies with an index of $\approx 2$,
  as expected in the test-particle limit, until the suppression of
  gamma rays due to $\pi^0$-decay kinematics is encountered and the
  curve is turning down. Note that already the extrapolation of the
  \hess\ spectrum is well below the EGRET upper limit from the
  position of \rxj, introduced in Sect.~\ref{subsec:spectra}. Taking
  into account non-linear effects would harden the gamma-ray spectrum
  even more.
  
  One should mention at this point that on the theory side other
  mechanisms to suppress contributions from low-energy (E around
  10~GeV) protons have been considered~\citep[e.g.,][]{Malkov}. When
  the particles upstream of the shock hit a dense target with a
  spatial gradient, such as a dense molecular cloud, the gamma-ray
  emission of low-energy protons might also be suppressed due to the
  energy dependence of the diffusion length. In a more general
  context, such mechanisms - an accelerator of finite lifetime
  interacting with a target at a distance where diffusion time scales
  are comparable to the source lifetime - have been studied by
  \citet{Aharonian96}. Such arguments, however, need to be reassessed
  when the exact location of the clouds with respect to the shock
  front is known.

  To calculate the energetics in a hadronic scenario, the mean target
  gas density available for gamma-ray production in the region of
  \rxj\ is a key question. The CO image shown in
  Fig.~\ref{fig:COImage}, left-hand side, reveals a hole in the CO
  emission and accordingly in the molecular hydrogen distribution in
  the central and eastern part of the SNR. In contrast, the TeV
  emission fills the whole region of the SNR (see
  Fig.~\ref{fig:Size200Image}). As is shown on the right-hand side of
  Fig.~\ref{fig:COImage}, there is no exact correlation between VHE
  gamma rays and CO intensity. From this one can conclude that in all
  likelihood cosmic rays do not penetrate the clouds uniformly. The
  bulk of the VHE gamma rays is not linked to the molecular clouds but
  must be due to interactions with a different target. Indeed, the
  rather good spherical shape of the remnant together with the fact
  that the X-ray and gamma-ray emission only varies by a factor of two
  to four across the remnant lends further support to the scenario
  that the SNR is running into a more or less uniform and probably
  low-density environment. Although it seems to be beginning to
  interact with the dense clouds to the west, the ones that are seen
  by NANTEN, VHE gamma rays are dominantly produced in cosmic-ray
  interactions with rather uniform ambient gas. One possible scenario
  is that the SNR is the result of a core-collapse supernova explosion
  that occurred into a wind bubble of a massive progenitor star. An
  SNR shock expanding into the bubble, with an ambient density much
  lower than that suggested by an average molecular cloud scenario,
  could explain the relative uniformity of the gamma-ray emission,
  compared to the large density variations in the clouds which likely
  surround the remnant (for a theoretical treatment of such
  configurations, see \citet{Berezhko2000}).

  The local target density is a crucial parameter in this
  scenario. With the NANTEN measurement of the void in the central
  part of the SNR one might constrain the local density in that
  region. The sensitivity of the final NANTEN data set as quoted in
  \citet{FukuiUchiyama} corresponds to a molecular column density of
  $8.3 \times 10^{19}~\mathrm{cm}^{-2}$ assuming the conventional
  conversion relation from CO intensity to $\mathrm{H}_2$ column
  density ($0.4~\mathrm{K}~\mathrm{km}~\mathrm{s}^{-1}$). Taking the
  diameter of the SNR as $\approx 20~{\mathrm{pc}}$ for
  $1~{\mathrm{kpc}}$ distance, one can deduce an upper limit on the
  molecular hydrogen column density of $\approx 2.6~\mathrm{cm}^{-3}$
  in parts of the SNR without detectable CO emission. The other
  existing constraint was inferred in~\citet{CassamXMM} from XMM data,
  based on the lack of thermal X-ray emission. By fitting the spectra
  with an absorbed power-law model and adding a thermal component, an
  upper limit on the mean gas temperature and, important here, the
  mean hydrogen number density of the ambient pre-shock medium of
  $0.02~\mathrm{cm}^{-3}$ was obtained. One should note, though, that
  this value is likely to be too low -- if the shocks are strongly
  modified by the accelerated particles, the shock heating is
  substantially reduced and the data would be consistent with higher
  densities.

  Assuming for now a mean target gas density of $n \approx
  1~\mathrm{cm^{-3}}$, uniformly spread throughout the remnant, in
  accordance with the NANTEN, but not the XMM limit, one can calculate
  the proton energetics implied by the gamma-ray flux measured from
  0.2 to 40~TeV. The total energy in accelerated protons from about
  $2$--$400$~TeV, required to provide the observed flux, can be
  estimated as $W^\mathrm{tot}_{p}(2$--$400~\mathrm{TeV}) \approx
  t_{pp \rightarrow \pi^0} \times L_\gamma (0.2$--$40~\mathrm{TeV})$,
  where $t_{pp \rightarrow \pi^0} \approx 4.5 \times 10^{15}
  (n/1~\mathrm{cm}^{-3})^{-1}~\mathrm{s}$ is the characteristic
  cooling time of protons through the $\pi^0$ production channel,
  $L_\gamma (0.2$--$40~\mathrm{TeV}) = 4\pi d^2 w_\gamma
  (0.2$--$40~\mathrm{TeV})$ is the luminosity of the source in gamma
  rays between 0.2 and 40~TeV, and $w_\gamma (0.2$--$40~\mathrm{TeV})$
  is the gamma-ray energy flux for the corresponding energy
  range. Assuming then that the proton spectrum with spectral index
  $\alpha \approx \Gamma$ continues down to 1~GeV, the total energy in
  protons can be estimated and compared to the total assumed
  mechanical explosion energy of the supernova of
  $10^{51}~\mathrm{erg}$. These calculations reveal very similar
  values for the three spectral shapes given in
  Fig.~\ref{fig:PlotHadrons} in the gamma-ray energy range between 0.2
  and 40~TeV: for the gamma-ray energy flux one obtains $w_\gamma
  (0.2$--$40~\mathrm{TeV}) \approx
  10^{-10}~\mathrm{erg}~\mathrm{cm}^{-2}~\mathrm{s}^{-1}$, the
  gamma-ray luminosity is $L_\gamma (0.2$--$40~\mathrm{TeV}) \approx
  10^{34}~\left(
  \frac{d}{1000~\mathrm{pc}}\right)^2~\mathrm{erg}~\mathrm{s}^{-1}$,
  and the corresponding energy content of protons is
  $W^\mathrm{tot}_{p}(2$--$400~\mathrm{TeV}) \approx 6 \times
  10^{49}~\left( \frac{d}{1000~\mathrm{pc}} \right)^2~\left(
  \frac{n}{1~\mathrm{cm}^{-3}}\right)^{-1}~\mathrm{erg}$. The
  resulting total energy in protons, after extrapolating the proton
  spectrum to 1~GeV and using $E_{51} \equiv 10^{51}~\left(
  \frac{d}{1000~\mathrm{pc}} \right)^2~\left(
  \frac{n}{1~\mathrm{cm}^{-3}}\right)^{-1}~\mathrm{erg}$, is then
  $W^{\mathrm{tot}}_p \approx 0.19 \times E_{51}$ for a power law with
  exponential cutoff, $W^{\mathrm{tot}}_p \approx 0.08 \times E_{51}$
  for a power law with energy dependent index, and $W^{\mathrm{tot}}_p
  \approx 0.26 \times E_{51}$ for a broken power law. These numbers
  are consistent with the notion of an SNR origin of Galactic cosmic
  rays involving the canonical $\approx 10\%$ conversion efficiency of
  the total supernova explosion energy. The \hess\ gamma-ray flux
  level is close to what was predicted in~\citet{DAV} from nearby
  young SNRs for ambient densities of $n \approx
  1~\mathrm{cm}^{-3}$. One should keep in mind though that the order
  of magnitude uncertainties in the measurements of the distance to
  the source $d$ and of the local gas density $n$ feed directly into
  these estimates.

  \subsection{Discussion and conclusions}\label{subsec:Discussion}
  The models and ideas presented in this section were aiming at
  exploring the possibilities available in explaining the observed VHE
  emission in purely electronic and purely hadronic scenarios. It is
  found that in the hadronic scenario, assuming gamma rays to stem
  from $\pi^0$ decays, the extrapolation of the \hess\ spectrum to
  lower gamma-ray energies leads to a picture that is consistent with
  the low-energy EGRET data. Furthermore, the spectral shape is well
  compatible with cosmic-ray acceleration theory. The energy
  requirements implied by the gamma-ray flux are in agreement with
  expectations from cosmic-ray acceleration in shell-type SNRs in our
  Galaxy, if one assumes a local target gas density of $n \approx
  1~\mathrm{cm}^{-3}$ and takes the currently preferred distance
  estimate of 1~kpc. Unfortunately both of these parameters are not
  very well measured. The distance estimate, which factors
  quadratically into the energetics calculation, has uncertainties in
  the order of at least 30\%. For the local target density there exist
  only upper limits, since from comparisons with CO data it turns out
  that gamma rays are most likely not exclusively linked to the dense
  molecular clouds surrounding the SNR. These clouds, however, obscure
  the measurement of the actual local target material available for
  gamma-ray production. Only towards the interior and the southeast of
  the SNR, where there is a hole in the molecular column density,
  is there hope to actually measure and constrain the
  density. Existing estimates in these regions are the NANTEN upper
  limit of $2.6~\mathrm{cm}^{-3}$, which does not cause any problem
  with the assumption made above, and the XMM upper limit of
  $0.02~\mathrm{cm}^{-3}$ which, if correct, would seriously challenge
  the idea of a hadronic scenario of gamma-ray production at least for
  this object.

  In the electronic scenario, on the other hand, the data are not
  easily reproduced taking only IC emission into account. The very low
  magnetic field of $\approx 10~\mu\mathrm{G}$, fixed by the ratio of
  synchrotron to IC flux, exceeds typical interstellar values only
  slightly and is difficult to reconcile with the paradigm of the
  diffusive shock acceleration of cosmic rays at supernova shock waves
  which predicts strong field amplifications in the region of the
  shock~\citep{LucekBell,BellLucek,Bell2004}. In the case of \rxj\ it
  was indeed considered possible by \citet{Voelk2005} that the
  magnetic field strength at the SNR shock front significantly exceeds
  typical interstellar values.

  Complete understanding of gamma-ray emission processes can only be
  achieved by taking a broadband approach and using all the available
  measurements in the different wavelength regimes. In
  Sect.~\ref{sec:Multiwavelength} the TeV data set was compared to
  X-ray, radio and CO emission measurements of the region surrounding
  \rxj. While there is no obvious resemblance with the radio image, it
  turns out that there is a striking spatial correlation between the
  ASCA X-ray and the \hess\ gamma-ray data. Most of the emission
  regions seem to exhibit exactly the same morphology in both
  wavelength regimes. At first sight this supports the idea that
  X-rays and gamma rays are produced by the same particle population,
  namely electrons. Assuming a constant magnetic field throughout the
  remnant (not the most likely configuration), the intensity (and
  spectrum) of both synchrotron and IC radiation trace the density
  (and the spectrum) of electrons, giving rise to the observed
  correlation. If the VHE gamma rays were due to non-thermal
  Bremsstrahlung of electrons, which is correlated with gas density,
  the observed correlation could be due to a magnetic field and gas
  density correlation. However, as can be seen from
  Fig.~\ref{fig:ElCase1}, Bremsstrahlung dominates over IC radiation
  only for very large values of $n_H > 100~\mathrm{cm}^{-3}$, which
  are not compatible with the CO measurements from the centre of the
  SNR, as mentioned above. But even given such a high density it is
  questionable if density, field strength and electron spectra can be
  fine-tuned such that the experimental results are approximately
  reproduced. Another difficulty for an electronic interpretation
  arises from the observation by~\citet{CassamXMM} that the X-ray
  spectra are steeper in the presumed shock front in the west, where
  the blast wave probably impacts the molecular cloud, than in the
  southeast, where the front propagates into a low density medium. It
  is very difficult to explain why the spectral shape in X-rays, but
  not in gamma rays, changes significantly in distinct regions of the
  shock, if they stem from the same particle population. If on the
  other hand gamma rays originate dominantly from nucleonic cosmic
  rays, a spatial correlation between X-rays and gamma rays is not
  automatically ensured either. There are two possible scenarios. The
  correlation could point to a common acceleration process
  accelerating both electrons and protons -- indeed expected in the
  theory of diffusive shock acceleration -- such that the spatial
  distributions are to first order the same and only differ because of
  the different loss processes. The second alternative is a correlated
  enhancement of magnetic field and local gas density.

  Another possibility of course is that the VHE gamma rays are a
  roughly equal mixture of two components, produced by both electrons
  and protons. However, this scenario seems unlikely since the
  energy-independent gamma-ray morphology and the absence of
  variations in spectral shape would again require fine-tuning of
  parameters like the magnetic field $B$ and the ambient density $n$.

  We conclude that the straightforward and simplest approaches in both
  scenarios lead to problems and one has difficulties in finding
  unequivocal evidence for either of them when using all the available
  broadband data. Nevertheless, the shape of the gamma-ray spectrum
  favours a hadronic scenario.

  \section{Summary}\label{sec:Summary}
  The VHE gamma-ray emission of \rxj\ was measured with unprecedented
  precision with \hess. The accuracy of the measurement is now
  approaching the level of X-ray measurements of this source, with the
  distinct advantage that \hess\ covers the whole SNR within its field
  of view. With the 2004 data, a close spatial correlation between
  X-rays and gamma rays was found. The overall gamma-ray energy
  spectrum was measured over more than two decades. There are
  indications for a deviation from a pure power-law spectrum. The data
  seem to be reasonably well described by a power law with an
  exponential cutoff and a power law with energy dependent photon
  index, as well as a broken power law. At the current stage further
  investigations about the shape of the spectrum at the highest
  energies accessible to \hess\ are hampered by the limited event
  statistics. The large data set has allowed for a spatially resolved
  spectral study. No significant variation in the gamma-ray spectral
  shape over the SNR is found. The flux varies by more than a factor
  of two across the SNR. The northern and western parts of the shell,
  where the SNR is believed to impact molecular clouds, are
  significantly brighter than the remaining parts.

  We presented broadband modelling ideas and discussed \rxj\ in terms
  of the available data from all wavelength bands including the \hess\
  gamma-ray signal. Two scenarios were addressed, one where gamma rays
  originate from electrons and one where they originate from
  protons. In both cases the large uncertainties on crucial parameters
  like the magnetic field strength and the effective ambient density,
  which are not directly accessible to measurements, hamper decisive
  conclusions. Nevertheless, the proton scenario is favoured because
  of the shape of the gamma-ray spectrum. From the theory side, the
  remaining challenge is the connection of the different particle
  species, VHE electrons and nuclei, in a consistent broadband model
  of \rxj. Experimentally, with the current gamma-ray data set, more
  precise measurements of the surrounding molecular clouds are clearly
  needed in order to link emission regions of VHE gamma rays to
  regions of known density.

  \begin{acknowledgements}
    The support of the Namibian authorities and of the University of
    Namibia in facilitating the construction and operation of
    H.E.S.S. is gratefully acknowledged, as is the support by the
    German Ministry for Education and Research (BMBF), the Max Planck
    Society, the French Ministry for Research, the CNRS-IN2P3 and the
    Astroparticle Interdisciplinary Programme of the CNRS, the
    U.K. Particle Physics and Astronomy Research Council (PPARC), the
    IPNP of the Charles University, the South African Department of
    Science and Technology and National Research Foundation, and by
    the University of Namibia. We appreciate the excellent work of the
    technical support staff in Berlin, Durham, Hamburg, Heidelberg,
    Palaiseau, Paris, Saclay, and in Namibia in the construction and
    operation of the equipment. We also thank Y.~Uchiyama for
    supplying the ASCA X-ray data and assisting with the comparisons
    with the \hess\ data, and Y.~Moriguchi and Y.~Fukui for supplying
    the NANTEN data.
  \end{acknowledgements}

  \bibliographystyle{aa}
  \bibliography{main}

\begin{thebibliography}{61}
\expandafter\ifx\csname natexlab\endcsname\relax\def\natexlab#1{#1}\fi

\bibitem[{{Aharonian} \& {Atoyan}(1996)}]{Aharonian96}
{Aharonian}, F.~A. \& {Atoyan}, A.~M. 1996, \aap, 309, 917

\bibitem[{{Aharonian} \& {Atoyan}(1999)}]{Aharonian99}
{Aharonian}, F.~A. \& {Atoyan}, A.~M. 1999, \aap, 351, 330

\bibitem[{{Aharonian} {et~al.}(1994){Aharonian}, {Drury}, \& {Voelk}}]{ADV}
{Aharonian}, F.~A., {Drury}, L.~O., \& {Voelk}, H.~J. 1994, \aap, 285, 645

\bibitem[{{Aharonian} {et~al.}(2004{\natexlab{a}})}]{HessCalib}
{Aharonian et al. \textit{(H.E.S.S. Collaboration)}} 2004{\natexlab{a}}, APh, 22, 109

\bibitem[{{Aharonian} {et~al.}(2004{\natexlab{b}})}]{Hess1713}
{Aharonian et al. \textit{(H.E.S.S. Collaboration)}} 2004{\natexlab{b}}, \nat, 432, 75

\bibitem[{{Aharonian} {et~al.}(2005{\natexlab{a}})}]{HessVelaJr}
{Aharonian et al. \textit{(H.E.S.S. Collaboration)}} 2005{\natexlab{a}}, \aap, 437, L7

\bibitem[{{Aharonian} {et~al.}(2005{\natexlab{b}})}]{HessPKS2155}
{Aharonian et al. \textit{(H.E.S.S. Collaboration)}} 2005{\natexlab{b}}, A\&A, 430,
  865

\bibitem[{{Aharonian} {et~al.}(2005{\natexlab{c}})}]{HessCrab}
{Aharonian et al. \textit{(H.E.S.S. Collaboration)}} 2005{\natexlab{c}}, in
  preparation

\bibitem[{{Aharonian} {et~al.}(2005{\natexlab{d}})}]{HessPlaneScan}
{Aharonian et al. \textit{(H.E.S.S. Collaboration)}} 2005{\natexlab{d}}, accepted for
  publication in \apj; astro-ph/0510397

\bibitem[{{Aschenbach}(1998)}]{AschenbachVelaJr}
{Aschenbach}, B. 1998, \nat, 396, 141

\bibitem[{{Atkins} {et~al.}(2004){Atkins}, {Benbow}, {Berley}, {Blaufuss},
  {Bussons}, {Coyne}, {DeYoung}, {Dingus}, {Dorfan}, {Ellsworth}, {Fleysher},
  {Fleysher}, {Gisler}, {Gonzalez}, {Goodman}, {Haines}, {Hays}, {Hoffman},
  {Kelley}, {Lansdell}, {Linnemann}, {McEnery}, {Miller}, {Mincer}, {Morales},
  {Nemethy}, {Noyes}, {Ryan}, {Samuelson}, {Shoup}, {Sinnis}, {Smith},
  {Sullivan}, {Williams}, {Westerhoff}, {Wilson}, {Xu}, \&
  {Yodh}}]{MilagroAllSky}
{Atkins}, R., {Benbow}, W., {Berley}, D., {et~al.} 2004, \apj, 608, 680

\bibitem[{{Bamba} {et~al.}(2005){Bamba}, {Yamazaki}, {Yoshida}, {Terasawa}, \&
  {Koyama}}]{Bamba2005}
{Bamba}, A., {Yamazaki}, R., {Yoshida}, T., {Terasawa}, T., \& {Koyama}, K.
  2005, \apj, 621, 793

\bibitem[{{Baring} {et~al.}(1999){Baring}, {Ellison}, {Reynolds}, {Grenier}, \&
  {Goret}}]{Baring99}
{Baring}, M.~G., {Ellison}, D.~C., {Reynolds}, S.~P., {Grenier}, I.~A., \&
  {Goret}, P. 1999, \apj, 513, 311

\bibitem[{{Bell}(2004)}]{Bell2004}
{Bell}, A.~R. 2004, \mnras, 353, 550

\bibitem[{{Bell} \& {Lucek}(2001)}]{BellLucek}
{Bell}, A.~R. \& {Lucek}, S.~G. 2001, \mnras, 321, 433

\bibitem[{{Berezhko} {et~al.}(2003){Berezhko}, {Ksenofontov}, \&
  {V{\"o}lk}}]{Berezhko2003}
{Berezhko}, E.~G., {Ksenofontov}, L.~T., \& {V{\"o}lk}, H.~J. 2003, \aap, 412,
  L11

\bibitem[{{Berezhko} \& {V{\" o}lk}(1997)}]{Berezhko97}
{Berezhko}, E.~G. \& {V{\" o}lk}, H.~J. 1997, APh, 7, 183

\bibitem[{{Berezhko} \& {V{\" o}lk}(2000)}]{Berezhko2000}
{Berezhko}, E.~G. \& {V{\" o}lk}, H.~J. 2000, \aap, 357, 283

\bibitem[{{Bernl{\" o}hr} {et~al.}(2003){Bernl{\" o}hr}, {Carrol}, {Cornils},
  {Elfahem}, {Espigat}, {Gillessen}, {Heinzelmann}, {Hermann}, {Hofmann},
  {Horns}, {Jung}, {Kankanyan}, {Katona}, {Khelifi}, {Krawczynski}, {Panter},
  {Punch}, {Rayner}, {Rowell}, {Tluczykont}, \& {van
  Staa}}]{BernloehrHessOptics}
{Bernl{\" o}hr}, K., {Carrol}, O., {Cornils}, R., {et~al.} 2003, APh, 20, 111

\bibitem[{{Blandford} \& {Eichler}(1987)}]{Blandford}
{Blandford}, R. \& {Eichler}, D. 1987, \physrep, 154, 1

\bibitem[{{Butt} {et~al.}(2002){Butt}, {Torres}, {Romero}, {Dame}, \&
  {Combi}}]{Butt}
{Butt}, Y.~M., {Torres}, D.~F., {Romero}, G.~E., {Dame}, T.~M., \& {Combi},
  J.~A. 2002, \nat, 418, 499

\bibitem[{{Cassam-Chena{\"i}} {et~al.}(2004){Cassam-Chena{\"i}},
  {Decourchelle}, {Ballet}, {Sauvageot}, {Dubner}, \& {Giacani}}]{CassamXMM}
{Cassam-Chena{\"i}}, G., {Decourchelle}, A., {Ballet}, J., {et~al.} 2004, \aap,
  427, 199

\bibitem[{{Cornils} {et~al.}(2003){Cornils}, {Gillessen}, {Jung}, {Hofmann},
  {Beilicke}, {Bernl{\" o}hr}, {Carrol}, {Elfahem}, {Heinzelmann}, {Hermann},
  {Horns}, {Kankanyan}, {Katona}, {Krawczynski}, {Panter}, {Rayner}, {Rowell},
  {Tluczykont}, \& {van Staa}}]{CornilsII}
{Cornils}, R., {Gillessen}, S., {Jung}, I., {et~al.} 2003, APh, 20, 129

\bibitem[{{Drury} {et~al.}(1994){Drury}, {Aharonian}, \& {V\"olk}}]{DAV}
{Drury}, L.~O., {Aharonian}, F.~A., \& {V\"olk}, H.~J. 1994, \aap, 287, 959

\bibitem[{{Ellison} {et~al.}(2001){Ellison}, {Slane}, \& {Gaensler}}]{Ellison}
{Ellison}, D.~C., {Slane}, P., \& {Gaensler}, B.~M. 2001, \apj, 563, 191

\bibitem[{{Enomoto} {et~al.}(2002){Enomoto}, {Tanimori}, {Naito}, {Yoshida},
  {Yanagita}, {Mori}, {Edwards}, {Asahara}, {Bicknell}, {Gunji}, {Hara},
  {Hara}, {Hayashi}, {Itoh}, {Kabuki}, {Kajino}, {Katagiri}, {Kataoka},
  {Kawachi}, {Kifune}, {Kubo}, {Kushida}, {Maeda}, {Maeshiro}, {Matsubara},
  {Mizumoto}, {Moriya}, {Muraishi}, {Muraki}, {Nakase}, {Nishijima}, {Ohishi},
  {Okumura}, {Patterson}, {Sakurazawa}, {Suzuki}, {Swaby}, {Takano}, {Takano},
  {Tokanai}, {Tsuchiya}, {Tsunoo}, {Uruma}, {Watanabe}, \&
  {Yoshikoshi}}]{CANGII}
{Enomoto}, R., {Tanimori}, T., {Naito}, T., {et~al.} 2002, \nat, 416, 823

\bibitem[{{Fukui} {et~al.}(2003){Fukui}, {Moriguchi}, {Tamura}, {Yamamoto},
  {Tawara}, {Mizuno}, {Onishi}, {Mizuno}, {Uchiyama}, {Hiraga}, {Takahashi},
  {Yamashita}, \& {Ikeuchi}}]{FukuiUchiyama}
{Fukui}, Y., {Moriguchi}, Y., {Tamura}, K., {et~al.} 2003, \pasj, 55, L61

\bibitem[{{Funk} {et~al.}(2004){Funk}, {Hermann}, {Hinton}, {Berge}, {Bernl{\"
  o}hr}, {Hofmann}, {Nayman}, {Toussenel}, \& {Vincent}}]{StefanHessTrigger}
{Funk}, S., {Hermann}, G., {Hinton}, J., {et~al.} 2004, APh, 22, 285

\bibitem[{{Ginzburg} \& {Syrovatskii}(1964)}]{Ginzburg}
{Ginzburg}, V.~L. \& {Syrovatskii}, S.~I. 1964, {The Origin of Cosmic Rays}
  (The Origin of Cosmic Rays, New York: Macmillan, 1964)

\bibitem[{{Hartman} {et~al.}(1999){Hartman}, {Bertsch}, {Bloom}, {Chen},
  {Deines-Jones}, {Esposito}, {Fichtel}, {Friedlander}, {Hunter}, {McDonald},
  {Sreekumar}, {Thompson}, {Jones}, {Lin}, {Michelson}, {Nolan}, {Tompkins},
  {Kanbach}, {Mayer-Hasselwander}, {M{\" u}cke}, {Pohl}, {Reimer}, {Kniffen},
  {Schneid}, {von Montigny}, {Mukherjee}, \& {Dingus}}]{Hartmann}
{Hartman}, R.~C., {Bertsch}, D.~L., {Bloom}, S.~D., {et~al.} 1999, \apjs, 123,
  79

\bibitem[{{Hillas}(1985)}]{Hillas}
{Hillas}, A.~M. 1985, in Proc. 19th ICRC, 445--448

\bibitem[{{Hinton}(2004)}]{JimHessStatus}
{Hinton, J.A. \textit{(H.E.S.S. Collaboration)}} 2004, New Astronomy Review, 48, 331

\bibitem[{{Hiraga}(2005)}]{HiragaPrivate}
{Hiraga}, J.~S. 2005, private communication

\bibitem[{{Hiraga} {et~al.}(2005){Hiraga}, {Uchiyama}, {Takahashi}, \&
  {Aharonian}}]{HiragaXMM}
{Hiraga}, J.~S., {Uchiyama}, Y., {Takahashi}, T., \& {Aharonian}, F.~A. 2005,
  \aap, 431, 953

\bibitem[{{Hofmann}(2003)}]{HofmannHessStatus}
{Hofmann, W. \textit{(H.E.S.S. Collaboration)}} 2003, in Proc. 28th ICRC, 2811

\bibitem[{{Jones} \& {Ellison}(1991)}]{Jones}
{Jones}, F.~C. \& {Ellison}, D.~C. 1991, \ssr, 58, 259

\bibitem[{{Katagiri} {et~al.}(2005){Katagiri}, {Enomoto}, {Ksenofontov},
  {Mori}, {Adachi}, {Asahara}, {Bicknell}, {Clay}, {Doi}, {Edwards}, {Gunji},
  {Hara}, {Hara}, {Hattori}, {Hayashi}, {Itoh}, {Kabuki}, {Kajino}, {Kawachi},
  {Kifune}, {Kiuchi}, {Kubo}, {Kurihara}, {Kurosaka}, {Kushida}, {Matsubara},
  {Miyashita}, {Mizumoto}, {Muraishi}, {Muraki}, {Naito}, {Nakamori}, {Nakase},
  {Nishida}, {Nishijima}, {Ohishi}, {Okumura}, {Patterson}, {Protheroe},
  {Sakamoto}, {Sakamoto}, {Swaby}, {Tanimori}, {Tanimura}, {Thornton},
  {Tsuchiya}, {Watanabe}, {Yamaoka}, {Yanagita}, {Yoshida}, \&
  {Yoshikoshi}}]{KatagiriVelaJr}
{Katagiri}, H., {Enomoto}, R., {Ksenofontov}, L.~T., {et~al.} 2005, \apjl, 619,
  L163

\bibitem[{{Koyama} {et~al.}(1997){Koyama}, {Kinugasa}, {Matsuzaki},
  {Nishiuchi}, {Sugizaki}, {Torii}, {Yamauchi}, \& {Aschenbach}}]{Koyama}
{Koyama}, K., {Kinugasa}, K., {Matsuzaki}, K., {et~al.} 1997, \pasj, 49, L7

\bibitem[{{Koyama} {et~al.}(1995){Koyama}, {Petre}, {Gotthelf}, {Hwang},
  {Matsuura}, {Ozaki}, \& {Holt}}]{Koyama95}
{Koyama}, K., {Petre}, R., {Gotthelf}, E.~V., {et~al.} 1995, \nat, 378, 255

\bibitem[{{Lazendic} {et~al.}(2004){Lazendic}, {Slane}, {Gaensler}, {Reynolds},
  {Plucinsky}, \& {Hughes}}]{Lazendic}
{Lazendic}, J.~S., {Slane}, P.~O., {Gaensler}, B.~M., {et~al.} 2004, \apj, 602,
  271

\bibitem[{{Lemoine-Goumard} \& {de Naurois}(2005)}]{ModelPaper}
{Lemoine-Goumard, M. \& de Naurois, M. \textit{(H.E.S.S. Collaboration)}} 2005, in AIP Conf. Proc. 745: High
  Energy Gamma-Ray Astronomy, 703--708

\bibitem[{{Lucek} \& {Bell}(2000)}]{LucekBell}
{Lucek}, S.~G. \& {Bell}, A.~R. 2000, \mnras, 314, 65

\bibitem[{{Malkov} {et~al.}(2005){Malkov}, {Diamond}, \& {Sagdeev}}]{Malkov}
{Malkov}, M.~A., {Diamond}, P.~H., \& {Sagdeev}, R.~Z. 2005, \apjl, 624, L37

\bibitem[{{Malkov} \& {O'C Drury}(2001)}]{MalkovDrury}
{Malkov}, M.~A. \& {O'C Drury}, L. 2001, Reports of Progress in Physics, 64,
  429

\bibitem[{{Mastichiadis} \& {de Jager}(1996)}]{Mastichiadis96}
{Mastichiadis}, A. \& {de Jager}, O.~C. 1996, \aap, 311, L5

\bibitem[{{Moriguchi} {et~al.}(2005){Moriguchi}, {Tamura}, {Tawara}, {Sasago},
  {Yamaoka}, {Onishi}, \& {Fukui}}]{Moriguchi}
{Moriguchi}, Y., {Tamura}, K., {Tawara}, Y., {et~al.} 2005, \apj, 631, 947

\bibitem[{{Muraishi} {et~al.}(2000){Muraishi}, {Tanimori}, {Yanagita},
  {Yoshida}, {Moriya}, {Kifune}, {Dazeley}, {Edwards}, {Gunji}, {Hara}, {Hara},
  {Kawachi}, {Kubo}, {Matsubara}, {Mizumoto}, {Mori}, {Muraki}, {Naito},
  {Nishijima}, {Patterson}, {Rowell}, {Sako}, {Sakurazawa}, {Susukita},
  {Tamura}, \& {Yoshikoshi}}]{CANGI}
{Muraishi}, H., {Tanimori}, T., {Yanagita}, S., {et~al.} 2000, \aap, 354, L57

\bibitem[{{Pannuti} {et~al.}(2003){Pannuti}, {Allen}, {Houck}, \&
  {Sturner}}]{Pannuti}
{Pannuti}, T.~G., {Allen}, G.~E., {Houck}, J.~C., \& {Sturner}, S.~J. 2003,
  \apj, 593, 377

\bibitem[{{Pfeffermann} \& {Aschenbach}(1996)}]{Pfeffermann}
{Pfeffermann}, E. \& {Aschenbach}, B. 1996, in Roentgenstrahlung from the
  Universe, 267--268

\bibitem[{{Piron} {et~al.}(2001){Piron}, {Djannati-Atai}, {Punch}, {Tavernet},
  {Barrau}, {Bazer-Bachi}, {Chounet}, {Debiais}, {Degrange}, {Dezalay},
  {Espigat}, {Fabre}, {Fleury}, {Fontaine}, {Goret}, {Gouiffes}, {Khelifi},
  {Malet}, {Masterson}, {Mohanty}, {Nuss}, {Renault}, {Rivoal}, {Rob}, \&
  {Vorobiov}}]{Piron}
{Piron}, F., {Djannati-Atai}, A., {Punch}, M., {et~al.} 2001, \aap, 374, 895

\bibitem[{{Reimer} \& {Pohl}(2002)}]{ReimerPohl}
{Reimer}, O. \& {Pohl}, M. 2002, \aap, 390, L43

\bibitem[{{Slane} {et~al.}(1999){Slane}, {Gaensler}, {Dame}, {Hughes},
  {Plucinsky}, \& {Green}}]{Slane}
{Slane}, P., {Gaensler}, B.~M., {Dame}, T.~M., {et~al.} 1999, \apj, 525, 357

\bibitem[{{Slane} {et~al.}(2001){Slane}, {Hughes}, {Edgar}, {Plucinsky},
  {Miyata}, {Tsunemi}, \& {Aschenbach}}]{SlaneVelaJr}
{Slane}, P., {Hughes}, J.~P., {Edgar}, R.~J., {et~al.} 2001, \apj, 548, 814

\bibitem[{{Tsuchiya} {et~al.}(2004){Tsuchiya}, {Enomoto}, {Ksenofontov},
  {Mori}, {Naito}, {Asahara}, {Bicknell}, {Clay}, {Doi}, {Edwards}, {Gunji},
  {Hara}, {Hara}, {Hattori}, {Hayashi}, {Itoh}, {Kabuki}, {Kajino}, {Katagiri},
  {Kawachi}, {Kifune}, {Kubo}, {Kurihara}, {Kurosaka}, {Kushida}, {Matsubara},
  {Miyashita}, {Mizumoto}, {Moro}, {Muraishi}, {Muraki}, {Nakase}, {Nishida},
  {Nishijima}, {Ohishi}, {Okumura}, {Patterson}, {Protheroe}, {Sakamoto},
  {Sakurazawa}, {Swaby}, {Tanimori}, {Tanimura}, {Thornton}, {Tokanai},
  {Uchida}, {Watanabe}, {Yamaoka}, {Yanagita}, {Yoshida}, \&
  {Yoshikoshi}}]{CangGalCen}
{Tsuchiya}, K., {Enomoto}, R., {Ksenofontov}, L.~T., {et~al.} 2004, \apjl, 606,
  L115

\bibitem[{{Uchiyama}(2005)}]{UchiyamaPrivate}
{Uchiyama}, Y. 2005, private communication

\bibitem[{{Uchiyama} {et~al.}(2003){Uchiyama}, {Aharonian}, \&
  {Takahashi}}]{UchiyamaChandra}
{Uchiyama}, Y., {Aharonian}, F.~A., \& {Takahashi}, T. 2003, \aap, 400, 567

\bibitem[{{Uchiyama} {et~al.}(2002){Uchiyama}, {Takahashi}, \&
  {Aharonian}}]{AscaI}
{Uchiyama}, Y., {Takahashi}, T., \& {Aharonian}, F.~A. 2002, \pasj, 54, L73

\bibitem[{{Vincent et al.}(2003)}]{PascalHessCamera}
{Vincent, P. et al. \textit{(H.E.S.S. Collaboration)}} 2003, in Proc. 28th ICRC, 2887

\bibitem[{{Vink} \& {Laming}(2003)}]{Vink2003}
{Vink}, J. \& {Laming}, J.~M. 2003, \apj, 584, 758

\bibitem[{{V{\"o}lk} {et~al.}(2005){V{\"o}lk}, {Berezhko}, \&
  {Ksenofontov}}]{Voelk2005}
{V{\"o}lk}, H.~J., {Berezhko}, E.~G., \& {Ksenofontov}, L.~T. 2005, \aap, 433,
  229

\bibitem[{{Wang} {et~al.}(1997){Wang}, {Qu}, \& {Chen}}]{Wang}
{Wang}, Z.~R., {Qu}, Q.-Y., \& {Chen}, Y. 1997, \aap, 318, L59

\end{thebibliography}
\end{document}